\documentclass[preprint,12pt,authoryear]{elsarticle}
\makeatletter
\def\ps@pprintTitle{%
 \let\@oddhead\@empty
 \let\@evenhead\@empty
 \def\@oddfoot{}%
 \let\@evenfoot\@oddfoot}
\makeatother
\usepackage[T1]{fontenc}
\usepackage{upgreek}
\usepackage{amssymb}
\usepackage{amsmath}
\usepackage{bm}
\usepackage{pifont}
\usepackage{geometry}
\usepackage{multirow}
\usepackage{tabularx}
\usepackage{array,booktabs}
\usepackage{colortbl}
\usepackage{lscape}
\usepackage{afterpage}

\newcolumntype{M}[1]{>{\centering\arraybackslash}m{#1}}
\newcolumntype{L}{>{\centering\arraybackslash}l}

\newcommand{\AlGaN}[2]{$\text{Al}_{#1}\text{Ga}_{#2}\text{N}$}
\newcommand{\AlGaNx}{\AlGaN{x}{1-x}}
\newcommand{\AlGaNSample}{\AlGaN{0.87}{0.13}}
\newcommand{\InGaNx}{$\text{In}_{x}\text{Ga}_{1-x}\text{N}$}

\newcommand{\sfrac}[2]{{}^#1/_#2}
\newcommand{\angstrom}{\textup{\AA}}
\newcommand{\degreem}{^{\circ}}
\newcommand{\degree}{$\degreem$\ }
\newcommand{\upDelta}{\Delta}
\newcommand{\vectorsym}[1]{\bm{#1}}

\newcommand{\aplusc}{$(\vectorsym{a}+\vectorsym{c})$}

\newcommand{\um}{$\upmu$m}
\newcommand{\umph}{\um\ hr\textsuperscript{-1}}

\begin{document}

\begin{frontmatter}

\title{PANIC: a 3D dislocation dynamics model for climb and glide in epitaxial films and heterostructures}

\author[Cambridge]{W. Y. Fu \corref{cor1}}
\ead{wyf22@cam.ac.uk}

\author[Cambridge]{C. J. Humphreys}
\author[Cambridge,Imperial]{M. A. Moram}

\cortext[cor1]{Corresponding author: Tel.: +44 (0)1223 334474.}

\address[Cambridge]{Dept. of Materials Science \& Metallurgy, University of Cambridge, 27 Charles Babbage Road, Cambridge CB3 0FS, United Kingdom}
\address[Imperial]{Dept. of Materials, Imperial College London, Exhibition Rd., London SW7 2AZ, United Kingdom}

\begin{abstract}
This paper presents PANIC, a 3D discrete mesoscale dislocation dynamics model which includes a fully quantitative treatment of both dislocation climb and dislocation glide, including climb driven by both osmotic and mechanical stresses and climb enabled by both bulk and pipe diffusion, including full elastic anisotropy for materials with hexagonal symmetry. Efficient calculations can be performed for epitaxial thin films, multilayers and device structures with free surfaces, including those with irregular geometries (e.g. islands). The model also includes the capability to simulate dislocation dynamics during the growth of the thin films or heterostructures. The model has been validated against experiment for thin films of GaN, AlN and AlGaN but is widely applicable to other material systems, both hexagonal and cubic.
\end{abstract}

\begin{keyword}
Dislocation dynamics
\sep Plasticity
\sep Thin films
\sep III-V Semiconductors
\sep GaN

\end{keyword}

\end{frontmatter}

\section{Introduction}
\label{sec:intro}

Dislocations are 1-dimensional line defects whose properties control the mechanical behaviour of many crystalline materials. Their existence was first proposed theoretically in 1934 by \citet{Taylor1934}, \citet{Orowan1934a, Orowan1934b, Orowan1934c} and \citet{Polanyi1934}, while subsequent studies by \citet{Hirsch1956}, \citet{Steeds1973}, \citet{Cottrell1963}, \citet{Hirth1982}, and many others described the properties of dislocations, focusing their effects on the plasticity of metals. More recently, research in this area has included atomistic simulations of dislocation core structures \citep{Belabbas2007,Cserti1992,Fang2000} and dislocation mobilities \citep{Olmsted2005,Weingarten2013}, as well as mesoscopic dislocation dynamics simulations used to model the microstructural development of metals containing high densities of dislocations and subjected to stresses \citep{vonBlanckenhagen2004,Bulatov2004,Devincre2011,Ghoniem2000,Groh2003,Mordehai2008,Schwarz1999,Weygand2002,Zbib2001}, and also multi-scale simulations spanning both atomistic and mesoscopic length scales \citep{Bulatov1998}. The majority of these simulations include dislocation glide, the rapid motion of dislocations within their slip plane in response to shear stresses acting across that plane. Dislocation climb can also occur, which is the non-conservative motion of dislocations perpendicular to their glide plane enabled by the diffusion of point defects towards or away from the dislocation core. However, the effects of dislocation climb are complex to calculate and greatly increase computational time. Earlier work includes dislocation climb as a simple 'drag factor' \citep{Cai2004}, and a bulk diffusion climb model has been properly introduced \citep{Ghoniem2000,Mordehai2008}. A full climb model has also been implemented recently in 2D by \citet{Ayas2014}, but this is computationally too expensive for 3D simulations.

The behaviour of dislocations in semiconductors has received considerable attention, particularly in GaN-based materials where heteroepitaxial thin films and devices contain dislocation densities from $10^7 - 10^9\ \text{cm}^{-3}$. Dislocations in III-nitride devices are known to reduce device lifetimes \citep{Furitsch2006,Mukai2006,Tapajna2011} and efficiencies \citep{Khan2008,Zhu2013}, as well as increasing leakage currents \citep{Chan2009,Kaun2011}. However, dislocations also affect the evolution of stresses during device growth. The high mismatch between the lattice parameters, thermal expansion coefficients and elastic constants of the III-nitride films and substrates mean that changes in epilayer composition or growth temperature are usually accompanied by changes in biaxial stresses. This is a major challenge for the growth of GaN-based devices on Si \citep{Dadgar2000,Haeberlen2010,Krost2002,Zhu2013} and for AlGaN-based devices for ultraviolet light emitters \citep{Amano1999,Han2001,McAleese2004}, in which cracking frequently occurs due to the biaxial stresses generated during or after growth. Dopants (such as Si) may also interact with dislocations, further affecting the evolution of stresses during device growth. Previous studies have shown that Si-doping is associated with an increase in tensile stress in the film, but it tends to ``pin'' dislocations, limiting climb \citep{Forghani2012,Moram2011a,Moram2011b}. There also appears to be a range of dislocation-mediated stress relaxation mechanisms in device structures: for instance, misfit dislocations are sometimes observed in \InGaNx/GaN structures \citep{Costa2005}, while inclined dislocations \citep{Chang2010} and ``staircase'' dislocations \citep{Cherns2008} dominate in \AlGaNx/GaN structures. These devices are based on either c-axis oriented epitaxial films, or (more recently) other 'nonpolar' and 'semipolar' orientated films. In the former case, dislocation climb is an important mechanism for dislocation mobility \citep{Fu2011,Moram2010} as the isotropic in-plane symmetry means that slip systems are rarely activated, despite the presence of high biaxial stresses. In the latter case, the loss of in-plane isotropy enables dislocation glide to occur in response to in-plane biaxial stress \citep{Hsu2011,Hsu2012}. This can lead to extended dislocation segments forming inside the active region of nonpolar or semipolar devices, further reducing device performance.

A large body of literature is devoted to the behaviour of dislocations within III-nitride semiconductor devices but it has not yet been possible to predict the behaviour of dislocations quantitatively. In this work, we present a dislocation dynamics model which enables prediction of the dislocation microstructure in III-nitride films and devices, named PANIC (Parallel ANIsotropic dislocation Climb and glide). This model includes a full treatment of both climb and glide, combining mesoscopic dislocation mechanics with parameters derived from atomistic simulations of dislocation behaviour, and takes into account the effects of the anisotropic elasticity of nitride crystals, the effects of thin film or 3D patterned geometries and the influence of multi-layer image stresses, temperature changes, and the biaxial stresses arising from mismatches in lattice parameters, elastic constants and thermal expansion coefficients between different epilayers and/or the substrate. In short, it includes every parameter that could reasonably be expected to affect dislocation movement in III-nitrides. The model can also be applied to thin film processes including annealing and epitaxial lateral overgrowth (ELOG). Here, we describe the model and validate it by comparison to experimental data on the dislocation microstructure in thin films of GaN and AlGaN.

\section{Methods}
\label{sec:method}
\begin{figure}[htbp]
\centering
\includegraphics[width=13cm]{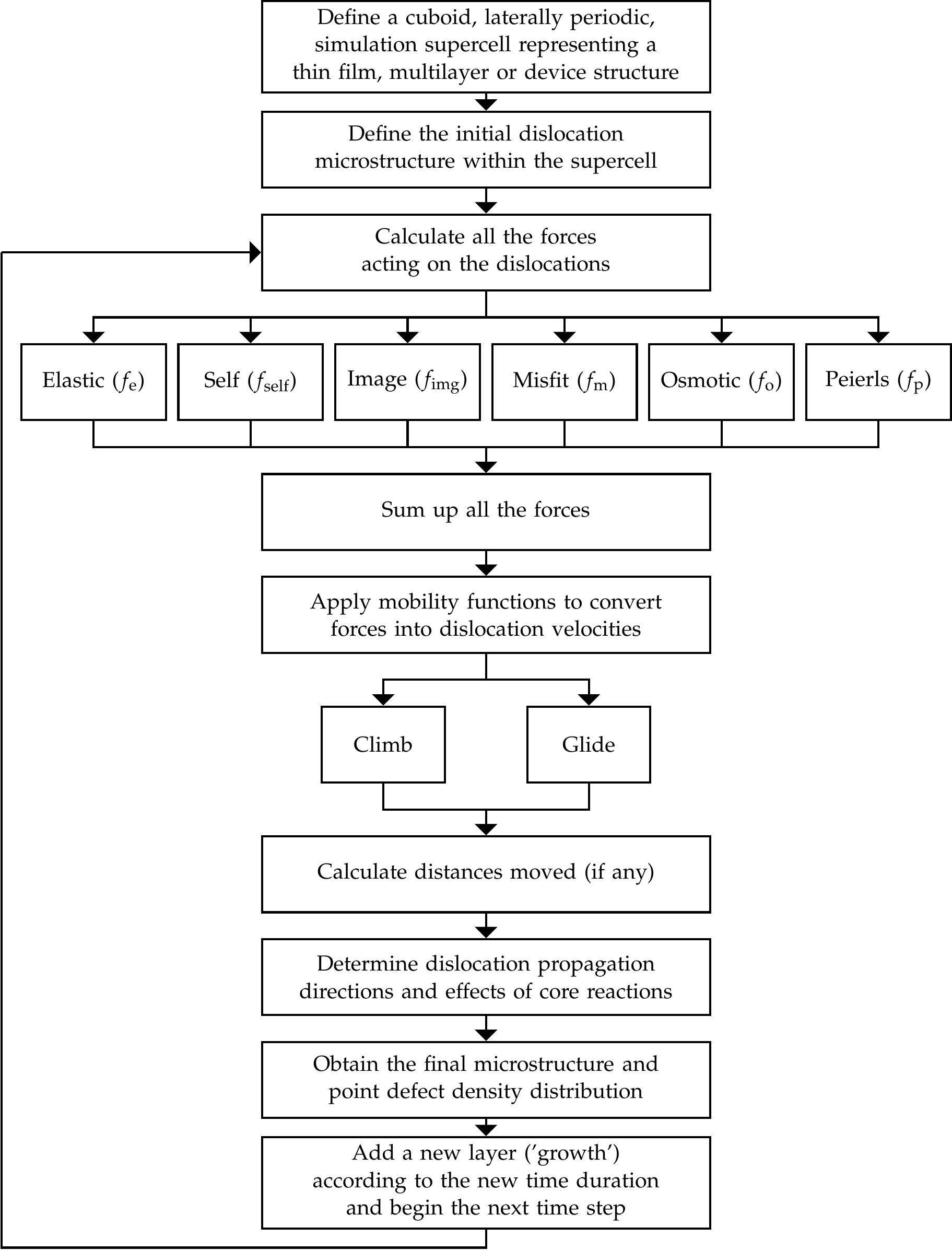}
\caption{Flow chart illustrating the approach used in this discrete dislocation dynamics model.}
	\label{fig:Fig1}
\end{figure}

An overview of the approach used in this model is given in Fig.~\ref{fig:Fig1} and is outlined in detail in the following section.

\subsection{Simulation supercell}
\begin{table}[htb]
\centering
\begin{tabular}{ p{6cm} p{6cm} }
\toprule
Sample parameters & Material parameters \\ \midrule
Dislocation density & Elasticity tensor $\vectorsym{C}$ \\
\rowcolor[gray]{0.9} Number of (threading) dislocations in the simulation cell &  Lattice parameters $a$ and $c$ \\
Ratio between $\vectorsym{c}$-type, $\vectorsym{a}$-type and \aplusc-type dislocations & Debye frequency $\omega_0$ \\
\rowcolor[gray]{0.9} Thin film, multilayer or device structure and geometry &  Dislocation core energy $E_\mathrm{core}$ \\
Crystallographic orientation of each layer & Temperature dependent thermal expansion function in the $\vectorsym{a}$ and $\vectorsym{c}$ directions \\
\rowcolor[gray]{0.9} Temperature &  Diffusion coefficients\textsuperscript{*} for bulk diffusion $D_\mathrm{B}$ \\
External pressure & Glide activation energy $Q_\mathrm{g}$ \\
\rowcolor[gray]{0.9} Growth rate &  Pipe diffusion climb activation energy $Q_\mathrm{p}$ \\
Growth direction & Atomic volume \\ \bottomrule
\multicolumn{2}{p{12cm}}{\footnotesize\textsuperscript{*}includes the vacancy or interstitial formation energy, the diffusion coefficient pre-exponential factor and the self-diffusion enthalpy}\\
\end{tabular}
\caption{Input parameters required for the dislocation dynamics simulation.}
\label{tab:Tab1}
\end{table}

The dislocation dynamics in III-nitride films and multilayers are modelled inside a cuboid anisotropic elastic space, which may include multiple layers of anisotropic elastic media. The dimensions $L_\mathrm{x}$ and $L_\mathrm{x}$ of the simulation space in the  $\mathbf{x}$- and $\mathbf{y}$- directions are kept equal to facilitate the use of periodic boundary conditions, which help to emulate the effect of a large specimen or wafer on the local simulated region. Periodic boundary condition is not implemented in the  $\mathbf{z}$- direction, however, so that the surface of the elastic space as well as the interface can be defined by one or more facet(s) which make up its geometry. Each layer in the simulation space is assumed to be crystalline and is allocated the properties summarised in Table~\ref{tab:Tab1}.

\begin{figure}
\centering
\includegraphics{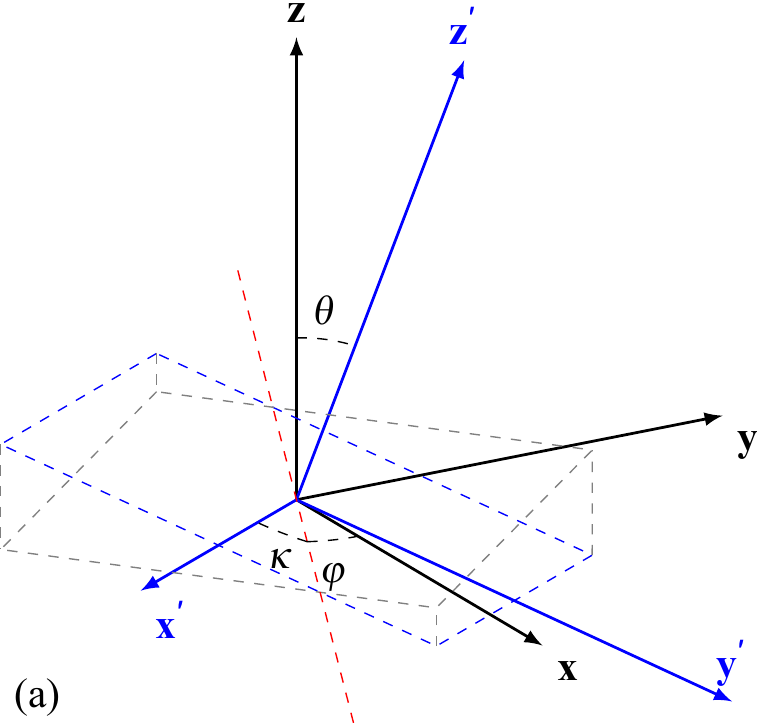}
\includegraphics{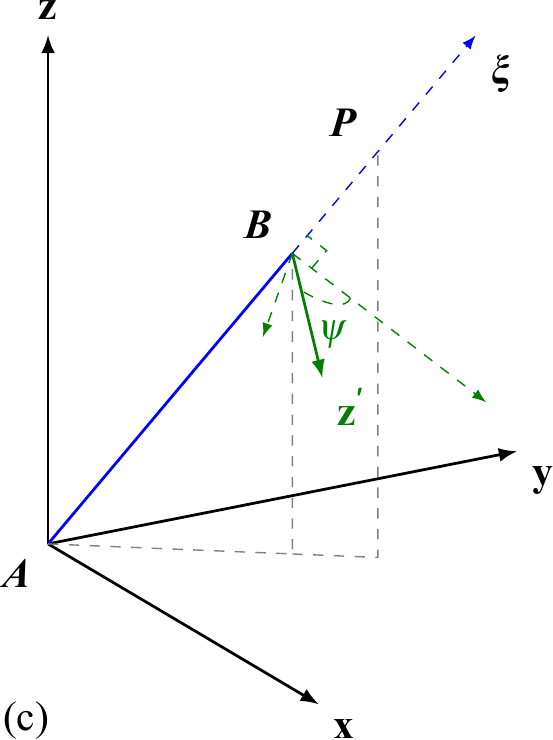}\\
\includegraphics{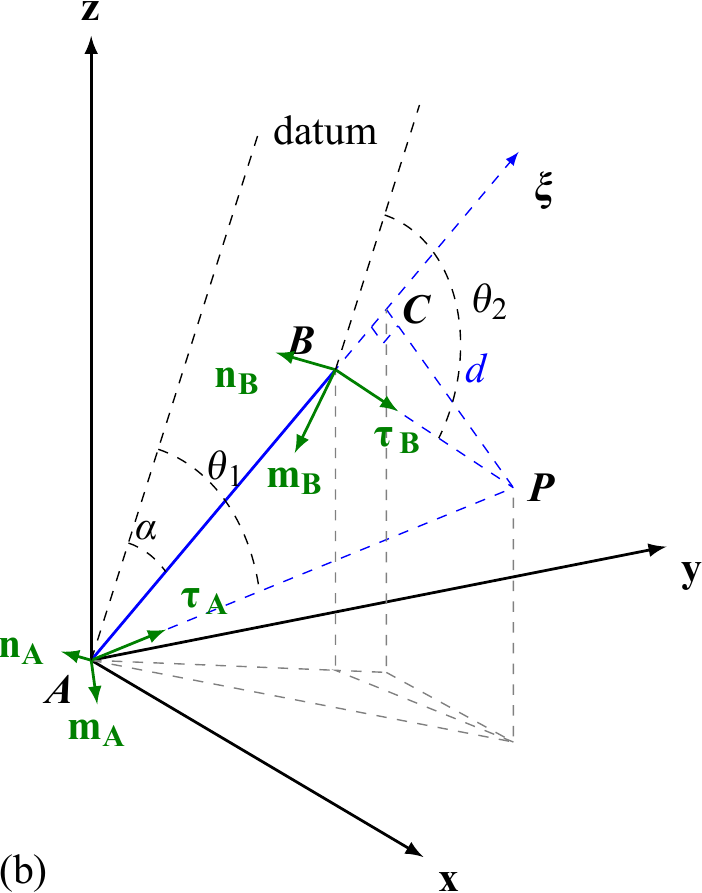}
\includegraphics{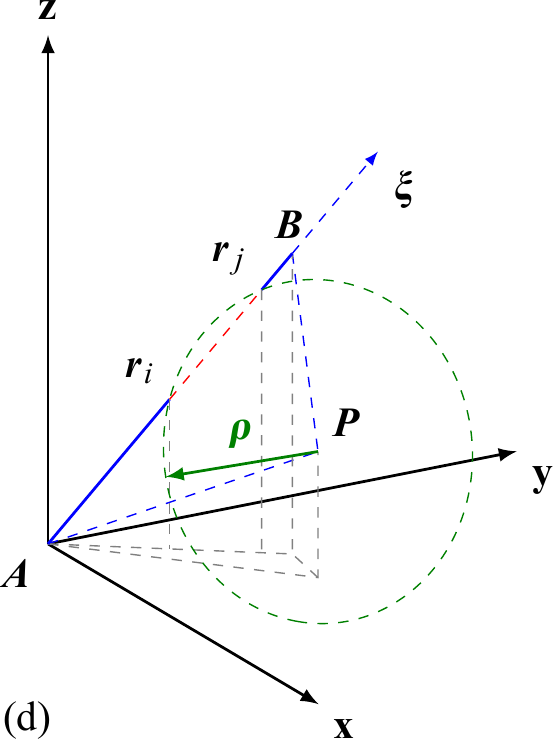}
\caption{(a) The relationship between the hexagonal and the Cartesian coordinate axes used in this model; (b) The geometric illustration of the Willis-Steeds-Lothe equation (Eqn.~\ref{eq:3}) and Brown's equation \citep{Brown1967}; and geometric setup for (c) collinear field points and (d) field points within the core radius $r_\mathrm{c}$ of a dislocation segment.}
	\label{fig:Fig2}
\end{figure}

The 4-index Miller-Bravais system is conventionally used to specify planes and vectors for hexagonal materials, however, it is more convenient to transform all 4-index vectors to Cartesian coordinates for geometric computation. This means that the simulation code can also be used for cubic materials. If the lattice vector $\left[2\bar{1}\bar{1}0\right]$ is parallel to the $\mathbf{x}$-axis and the lattice vector $\left[0001\right]$ is parallel to the $\mathbf{z}$-axis, then any lattice vector $\left[uvtw\right]$ or reciprocal lattice vector $\left(hkil\right)$ of a hexagonal lattice system can be converted to the Cartesian coordinate system $(x, y, z)$ using the following relations, where $a$ and $c$ are the lattice parameters of the hexagonal crystal:
\begin{equation}
\left[uvtw\right] = \frac{3}{2}ua \mathbf{x}+\frac{\sqrt{3}}{2}(u+2v)a \mathbf{y}+wc \mathbf{z}
\label{eq:1}
\end{equation}
\begin{equation}
\left(hkil\right) = \frac{h}{a} \mathbf{x}+\frac{h+2k}{\sqrt{3}a} \mathbf{y}+\frac{l}{c} \mathbf{z}
\label{eq:2}
\end{equation}
To account for semi-polar and non-polar film orientations, where the vector $[0001]$ is no longer aligned with the $\mathbf{z}$-axis, and for some films which may be deposited in a slightly different planar orientation with respect to the substrate, a coordinate orientation transformation is also implemented as described by \citet{Hirth1982}, by applying a transformation matrix $\vectorsym{T}$ to the specific coordinate $\vectorsym{r}$, with respect to the three Eulerian angles $\theta$, $\phi$ and $\kappa$ for transformation (see Fig.~\ref{fig:Fig2}(a)). The stress $\vectorsym{\sigma}$, strain $\vectorsym{\epsilon}$ and stiffness tensor $\vectorsym{C}$ can thus be transformed using the transformation matrix accordingly \citep{Hirth1982}.

\subsection{Dislocation modelling}
Pure dislocations in the III-nitrides can have any of the three possible Burgers vectors: $\sfrac{1}{3}\left<11\bar{2}0\right>$  ($\vectorsym{a}$-type), $\left<0001\right>$ ($\vectorsym{c}$-type) and $\sfrac{1}{3}\left<11\bar{2}3\right>$ (\aplusc-type). Dislocations are first generated using positions and Burgers vectors assigned randomly using a Mersenne Twister algorithm \citep{Matsumoto1998}. For a typical simulation, each dislocation has a line vector $\vectorsym{\xi}$ which is the same as the normal vector of the substrate plane, although manual setup of dislocation positions and orientations is also possible. The generated dislocations are then divided into small segments defined by positions of the two end nodes, so that segments with the same node are connected together. The minimum length of the dislocation segment is limited by the core radius of the dislocation and the magnitude of the error in the calculated dislocation movement arising from the time integration, while the maximum length is limited by the error related to the angle between two straight segments, which is proportional to the deviation from a long straight segment compared to a curved segment. This discretization allows freedom of movement of the dislocation line in each simulation step. At each time step, the segmentation of the dislocation is adaptively recalculated according to the above criteria and the angle between two segments, thus minimizing the computational power needed, while maintaining an acceptable spatial resolution.

\subsection{Forces affecting dislocation movement}
In this model, we consider and model the driving forces for the evolution of the dislocation microstructure existing in thin films under realistic growth conditions. At equally spaced points on each dislocation segment, the forces that are important to both glide and climb in multi-layered structures are calculated, including the elastic force, self-force, image force, misfit force, osmotic force and Peierls force. By integrating the points (using Simpson's rule) on a node's connected segments with respect to a shape function, the segment now acts like it is pinned to the two nodes at its end points. The total integrated force can then be used to calculate the actual dislocation movement velocity using an adaptive 4\textsuperscript{th}-5\textsuperscript{th} order Adam-Bashforth-Moulton predictor corrector method \citep{Mathews2006}, with the first 3 time steps being initialized using the Euler-trapezoid method.

\subsubsection{Elastic forces}
The elastic interactions within and between dislocations are modelled by the cumulative effect of the computed stress fields from each dislocation on individual dislocation line segment. The summed stresses lead to an expression for the elastic force $f_\mathrm{e}$ per unit length $L$.

The initial method to calculate the stress field of a straight dislocation segment of finite length is proposed by \citet{Brown1967} as a proof to the theorem of \citet{Lothe1967}, who expressed dislocation forces in terms of energy factors. However, this method causes singularity in collinear cases, thus we adopted the non-singular expression of \citet{Willis1970} , as described by \citet{Yin2010}. This Willis-Steeds-Lothe formula describes the stress tensor field on a dislocation segment   in an infinite anisotropic elastic space (using the Einstein summation convention):
\begin{equation}
\sigma_{ij}(\vectorsym{r}) = \frac{1}{4 \mathrm{\uppi} d} \upepsilon_{qln} b_p C_{ijkl} C_{pqms} \xi_n \Big\{ -m_s Q_{km} + n_s \left[ (nn)^{-1} \cdot (nm) \cdot Q \right]_{km} + n_s \left[ (nn)^{-1} \cdot S^{\mathrm{T}} \right]_{km} \Big\} \Big|^{\vectorsym{BP}}_{\vectorsym{AP}}
\label{eq:3}
\end{equation}
where $\vectorsym{\sigma}$ is the stress tensor; $\boldsymbol{\upepsilon}$ is the permutation tensor; $d$ is the distance $|\vectorsym{CP}|$ from the dislocation line vector to the field point; $C_{ijkl}$ is the component of the stiffness tensor $\vectorsym{C}(\vectorsym{r})$; $\vectorsym{b}$ is the Burgers vector; $\vectorsym{\xi}$ is the dislocation line vector; $\mathbf{m}$, $\mathbf{n}$ and $\boldsymbol{\uptau}$ form an orthogonal basis as illustrated in Fig.~\ref{fig:Fig2}(b), with the field point $\vectorsym{P}$ (at coordinate $\vectorsym{r}$) and the line vector $\vectorsym{\xi}$ staying on the same plane and $\mathbf{n}$ being this plane's normal vector. The term $(pq)_{jk}$, with vectors $\vectorsym{p}$ and $\vectorsym{q}$, is the Christoffel stiffness tensor given by $p_{i}C_{ijkl}q_{j}$, and $(pq)^{-1}_{jk}$ is its inverse. The tensors $\vectorsym{Q}$ and $\vectorsym{S}^{\mathrm{T}}$ are as determined by \citet{Asaro1974}.

To model collinear dislocation segments, the expression becomes (see Fig.~\ref{fig:Fig2}(c)) \citep{Yin2010}:
\begin{equation}
\sigma_{kl}(\vectorsym{r}) = \left( \frac{1}{|\vectorsym{BP}|} - \frac{1}{|\vectorsym{AP}|} \right) g_{kl}(\vectorsym{\xi})
\label{eq:4}
\end{equation}
where
\begin{equation}
g_{kl}(\vectorsym{\xi}) = \frac{1}{8 {\uppi} ^2} C_{klip} \upepsilon_{pjw} b_m C_{wmrs} \xi_j \int^{2\pi}_0 \big\{ \xi_s (z' z')^{-1}_{ir} - z'_s [ (z' z')^{-1} [(\xi z')+(z' \xi)] (z' z')^{-1}]_{ir} \big\} \mathrm{d} \psi
\label{eq:5}
\end{equation}
The resulting force $\vectorsym{f}$ on the original dislocation line segment can be calculated from the stress $\bm{\sigma}$ using the Peach-K\"{o}hler equation $\vectorsym{f} = \vectorsym{b} \cdot \bm{\sigma} \times \vectorsym{\xi}$.

\subsubsection{Self-forces}
The self-force of a dislocation is the force exerted on a dislocation by itself, which tends to minimise the length of the dislocation and which contains two contributions, the elastic and the core self-force. 
The elastic self-force is calculated using the same formula as for the collinear dislocation segments (Eqn.~\ref{eq:4}), but using an inner cutoff radius \citep{Yin2010} (see Fig.~\ref{fig:Fig2}(d)). The inner cut-off radius is applied to avoid the singularity arising from classical continuum dislocation theory \citep{Cai2006} which would lead to a divergence of stress fields computed when the field point is close to the dislocation core. It is usually chosen to be the dislocation core radius. This assumption appears to be reasonable for III-nitrides, as previous critical thickness calculations by \citet{Holec2008} show a good match with experimental data using the core radius as the inner cutoff radius. 

The self-force from the core region can be divided into two components: a longitudinal self-force and a torsional self-force \citep{Fitzgerald2010}. The former force will act to reduce the dislocation segment length (thereby minimising its total energy), while the latter force will tend to rotate the segment line direction to a more energetically favourable direction. In an anisotropic elastic space, the core longitudinal force   for each dislocation segment is \citep{Fitzgerald2010}: 
\begin{equation}
\vectorsym{f}^{\mathrm{L}}_{\mathrm{c}} = -\frac{4 \uppi E_{\mathrm{core}}}{\bar\mu |\vectorsym{b}|^2} (\vectorsym{b} \cdot \bm{B} \cdot \vectorsym{b} ) \vectorsym{\xi}
\label{eq:6}
\end{equation}
where $\bar{\mu}$ is the anisotropic shear modulus; $E_{\mathrm{core}}$ is the core energy; and $\vectorsym{B}$ can be calculated by
\begin{equation}
B_{ij} = \frac{1}{8 \uppi^2} \int^{2 \uppi}_0 \left\{ (mm)_{ij} - (mn)_{ik}(nn)^{-1}_{kl}(nm)_{lj} \right\} \mathrm{d} \psi
\label{eq:7}
\end{equation}
And the core torsional force, which tends to rotate the dislocation segments to lower energy configuration, would be \citep{Fitzgerald2010}
\begin{equation}
\vectorsym{f}^{\mathrm{T}}_{\mathrm{c}} = -\frac{4 \uppi E_{\mathrm{core}}}{\bar\mu |\vectorsym{b}|^2} \left[ \left(\vectorsym{b} \cdot \frac{{\partial} \bm{B}}{{\partial} \phi} \cdot \vectorsym{b} \right) \mathbf{\mathrm{e}}_{\boldsymbol{\upphi}} + \left(\vectorsym{b} \cdot \frac{{\partial} \bm{B}}{{\partial} \theta} \cdot \vectorsym{b} \right) \mathbf{\mathrm{e}}_{\boldsymbol{\uptheta}} \right] 
\label{eq:8}
\end{equation}
where $\mathbf{\mathrm{e}}_{\boldsymbol{\upphi}}$ and $\mathbf{\mathrm{e}}_{\boldsymbol{\uptheta}}$ are the base vectors as in a spherical coordinate system.

\subsubsection{Image forces}
The Willis-Steeds-Lothe formula describes the stress fields of dislocations in an infinite anisotropic elastic space, but surfaces and interfaces play a significant role in semiconductor thin films by imposing 'image forces' on the dislocations \citep{Hirth1982}. Expressions already exist for the stress field of a semi-infinite straight dislocation with arbitrary Burgers vector and angle of incidence, terminating at the free surface of an elastically anisotropic semi-infinite solid \citep{Head1953a,Head1953b,Yoffe1961,Lothe1982}. However, to calculate image forces for an arbitrary dislocation microstructure with multiple interfaces, we follow the superposition approach \citep{Fivel1998,VanderGiessen1995,Yasin2001}, in which the calculation is decomposed into two separate elastic problems: (1) the problem of interacting dislocations in a homogeneous infinite elastic solid, and (2) a dislocation-free version of the original problem. To account for dislocation-boundary (surface) intersection, the dislocation segment that terminates on the surface is extended up to 1000 times the original segment, producing the 'augmented dislocation configuration' \citep{Deng2008,Weinberger2009}.

We have therefore used a hybrid finite element method combining the method of superposition of multiple subdomains of \citet{Zbib2002} with the virtual segment approach of \citet{Weinberger2009}. This provides the flexibility to simulate arbitrary interfaces and surfaces and also has an acceptable accuracy provided that the finite element 'mesh' is fine enough near the dislocation-surface interception point.

To implement this approach, each anisotropic elastic layer is treated separately to satisfy a traction-matching condition at the interfaces. The traction force resulting from the image stress $\tilde{T}_{i}$ at the interface between the $i^\mathrm{th}$ and $(i+1)^\mathrm{th}$ layer is estimated according to:
\begin{equation}
\tilde{\vectorsym{T}}_{i} = \bar{\gamma} \vectorsym{T}^{\infty}_i
\label{eq:9}
\end{equation}
$\vectorsym{T} = \bm{\sigma} \cdot \vectorsym{n}$; $\vectorsym{T}^{\infty}_i$ is the traction force on the interface in an infinite anisotropic elastic space such that the resulting traction force $\vectorsym{T}_i = \vectorsym{T}^{\infty}_i + \tilde{\vectorsym{T}}_{i}$; and $\bar{\gamma} = \dfrac{\bar{\mu}_{i+1}-\bar{\mu}_i}{\bar{\mu}_{i+1}+\bar{\mu}_i}$. The resulting traction force at the surface with normal $\vectorsym{n}$ would be reduced to zero for a traction-free condition with $\bar{\gamma} = -1$, that is, the image traction force would be a 'reverse' of that in the infinite anisotropic elastic space. It is also obvious that there would be no imaging traction force when layer $i$ and layer $i+1$ are the same, causing $\bar{\gamma} = 0$.

\begin{figure}
\centering
\includegraphics{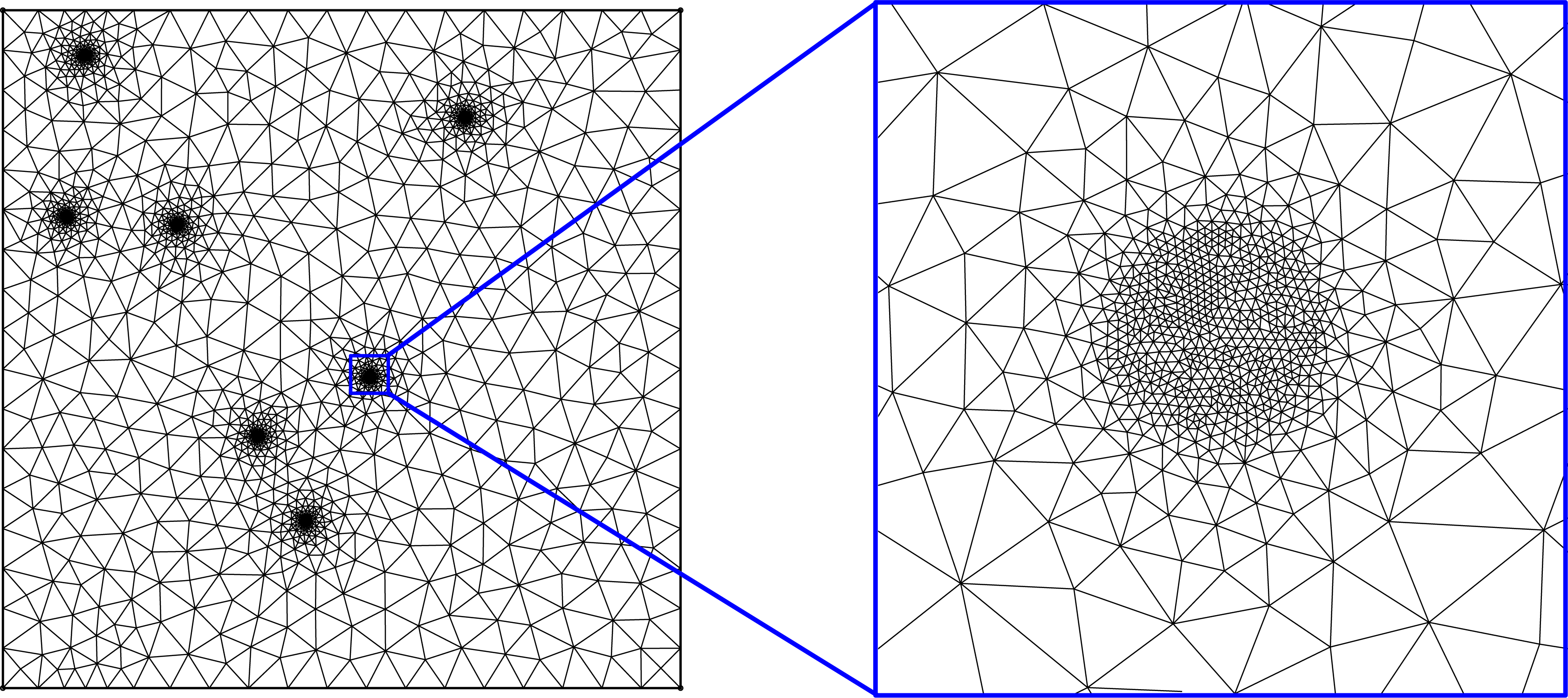}
\caption{Adaptive mesh on a film surface, showing elements near the dislocation segment which have a dimension of about 2 \angstrom (i.e. atomic scale). The larger area shown has dimensions of 800 nm $\times$ 800 nm.}
\label{fig:Fig3}
\end{figure}

Setting $\tilde{\vectorsym{T}}_{i}$ as the boundary condition of the finite element calculation, the image stress tensor field can be solved using an external 3D finite element mesh generator \emph{gmsh} \citep{Geuzaine2009} and solver \emph{getdp} \citep{Dular1998}. Since the image force is most intense near surface and interfaces, the mesh is generated such that it is adaptively meshed finely where the dislocation segments intercept with the interfaces and surfaces (Fig.~\ref{fig:Fig3}).
The image stress, when summed with the stress field in infinite anisotropic elastic space as calculated using the Willis-Steeds-Lothe equation, can satisfy both the traction-matching condition at the interfaces and the traction-free condition at the surface. The Willis-Steeds-Lothe equation (originally formulated for an infinite anisotropic space) is thus adapted to compute the stress field of a dislocation in a finite anisotropic space.

\subsubsection{Misfit forces}
The misfit force arises from the elastic stress generated by the lattice mismatch between film and substrate. To take into account the effect of the stress tensor field from other factors on the misfit force, the lattice parameters on each layer are re-calculated every time step from the total stress field excluding the misfit force itself. In this case, the misfit strain $\epsilon_m(t,T)$ on any layer would be
\begin{equation}
\epsilon_{m}(t,T) = \frac{a'_{\mathrm{f}}(t,T) - a_{\mathrm{f}}(T)}{a_{\mathrm{f}}(T)}
\label{eq:10}
\end{equation}
where $a'_{\mathrm{f}}(t,T)$ is the recalculated lattice parameter at time $t$ and temperature $T$; and $a_f(T)$ is the thermally expanded lattice parameters for the film. Note that Equation~\ref{eq:10} follows the convention that compressive strain is negative, and vice versa. The thermal expansion coefficients $\alpha(T)$ are determined using the Reeber model for GaN and AlN \citep{Reeber2000} and the Suzuki model for InN \citep{Wang2001}, respectively.
 
Assuming the lattice parameter of the film is lattice-matched to the substrate at time $t = 0$, i.e., $a'_{\mathrm{f}}(0,T) = a_{\mathrm{s}}(T)$, then at this time this equation reduces to the conventionally defined $\epsilon_m = \frac{a_{\mathrm{s}} - a_{\mathrm{f}}}{a_{\mathrm{f}}}$ \citep{Freund2003}, and the re-calculated lattice parameter at any time $t$ is as follows, where $\epsilon_{\parallel}$ is the equivalent strain (as calculated from other driving forces) parallel to the direction of  $a$ calculated from the strain-displacement equation:
\begin{equation}
a'_{\mathrm{f}}(t,T) = (1+\epsilon_{\parallel}) a_{\mathrm{s}}(T)
\label{eq:11}
\end{equation}
If the film does not have the same crystallographic orientation as the substrate, then the lattice parameters   in Eqn.~\ref{eq:10} and Eqn.~\ref{eq:11} can be replaced by the atomic plane spacing   so that the lattice parameters   and   can be solved in a system of parametric equations. So in this case, $d'_{\mathrm{f1}}(0,T) = d_{\mathrm{s1}}(T)$, $d'_{\mathrm{f2}}(0,T) = d_{\mathrm{s2}}(T)$, then the following applies, where $d_{\mathrm{f1}}(t,T) = d_{h_1 k_1 k_1}$ and $d_{\mathrm{f2}}(t,T) = d_{h_2 k_2 l_2}$:
\begin{equation}
a_{\mathrm{f}}^2 = \frac{4}{3} \frac{d_{\mathrm{s1}}^2 d_{\mathrm{s2}}^2 [l_2^2 ( h_1^2 + h_1 k_1 + k_1^2 ) - l_1^2 ( h_2^2 + h_2 k_2 + k_2^2 )]}{d_{\mathrm{s2}}^2 l_2^2 - d_{\mathrm{s1}}^2 l_1^2}
\label{eq:12}
\end{equation}
\begin{equation}
c_{\mathrm{f}}^2 = \frac{d_{\mathrm{s1}}^2 d_{\mathrm{s2}}^2 [l_2^2 ( h_1^2 + h_1 k_1 + k_1^2 ) - l_1^2 ( h_2^2 + h_2 k_2 + k_2^2 )]}{d_{\mathrm{s1}}^2 (h_1^2 + h_1 k_1 + k_1^2) - d_{\mathrm{s2}}^2 (h_2^2 + h_2 k_2 + k_2^2 )}
\label{eq:13}
\end{equation}
The misfit stress on a layer can thus be computed by applying a fixed displacement calculated from the corresponding misfit strain on its side wall in a finite element calculation. 

\subsubsection{Thermal forces}
External biaxial stress mainly originates from the different thermal expansion coefficients and growth temperatures of different layers. The thermal expansion coefficients of materials were included in the misfit force as described in the previous section. However, thermal mismatch stress can be incurred by the difference in thermal expansions and growth temperatures, in addition to the thermal mismatch arising from the lattice parameters due to difference in thermal expansion coefficient. The thermal mismatch force in this section simulates thermal effect from layers grown below the simulation environment. Using the thermal expansion function from \citet{Reeber2000} and \citet{Wang2001}, the thermal expansion vector, representing the thermal expansion in 3D, for a $\vectorsym{c}$-plane layer, would be
\begin{equation}
\vectorsym{\alpha}\left(T,T_\mathrm{g}\right) = \left[\alpha_\mathrm{a}\left(T,T_\mathrm{g}\right), \alpha_\mathrm{a}\left(T,T_\mathrm{g}\right), \alpha_\mathrm{c}\left(T,T_\mathrm{g}\right)\right]
\label{eq:14}
\end{equation}
where $T_\mathrm{g}$ is the growth temperature of the layer, $T$ is the growth temperature during the current time step.

So for a layer with an arbitrary orientation, by rotating the base vectors $\mathbf{x}$, $\mathbf{y}$ and $\mathbf{z}$ to $\vectorsym{c}$-plane from that orientation, the corresponding in-plane thermal expansion vector would be:
\begin{equation}
\vectorsym{\alpha}_i\left(T,T_\mathrm{g}\right) = \vectorsym{\alpha}\left(T,T_\mathrm{g}\right) \left[ \vectorsym{x}_\mathrm{c}^{\mathrm{T}}, \vectorsym{y}_\mathrm{c}^{\mathrm{T}}, 0 \right]
\label{eq:15}
\end{equation}
where $\vectorsym{x}_\mathrm{c}$ and $\vectorsym{y}_\mathrm{c}$ are the rotated base vectors. Here we assume that the thermal mismatch is a biaxial stress (acting only in planar direction), thus the expansion vector in the vertical ($\vectorsym{\mathrm{z}}$-) direction is zeroed. The differential thermal expansion vector can then be calculated from:
\begin{equation}
\upDelta \vectorsym{\alpha}_{i}\left(T,T_{\mathrm{g},i}\right) = \vectorsym{\alpha}_{i}\left(T,T_{\mathrm{g},i}\right) - \vectorsym{\alpha}_{(i-1)}\left(T,T_{\mathrm{g},(i-1)}\right)
\label{eq:16}
\end{equation}
with $\upDelta \alpha_{0} = 0$. Unlike the misfit force, the effect of the thermal force, which would only be dependent on the in-plane thermal mismatch, should be cumulative starting from the substrate, since each layer is grown on top of a thermally expanded previous layer. So the resultant thermal strain on the whole structure, assuming no bending, would be
\begin{equation}
\left[ \epsilon_{\mathrm{T}}(t,T) \right]_j = \sum_{i=0}^{n(t)} \upDelta \alpha_{ij}\left(T,T_{\mathrm{g},i}\right)
\label{eq:17}
\end{equation}
where $n(t)$ is the total number of layers at time $t$.

The thermal stress and thermal force can then be computed by using a finite element method similar to that for the mismatch stress, but with the fixed displacement calculated from the thermal strain instead.

\subsubsection{Osmotic forces}
Osmotic forces affect dislocation climb and are generated when a non-equilibrium concentration of point defects arises in the material. This creates a driving force for dislocations to restore equilibrium by the absorption or emission of point defects from the dislocation core (resulting in dislocation climb). The magnitude of the osmotic force is therefore controlled by point defect diffusion, and only contributes to dislocation climb. The point defect concentration field $c_\mathrm{v}$ are initialised by assuming that they are initially in equilibrium with dislocations. It is then subjected to bulk diffusion as the dislocations move during the simulation \citep{Hirth1982}:
\begin{equation}
\frac{{\partial}c_\mathrm{v}}{{\partial}t} = D\upDelta c_\mathrm{v}
\label{eq:18}
\end{equation}
where $D$ is the diffusivity. The diffusion of the vacancy concentration is implemented with a finite difference method (FDM) using a 3D forward time, centered space (FTCS) scheme with adaptive time stepping. Section 2.4.2 then describes how the bulk vacancy concentration field is used to approximate climb assisted by pipe diffusion, as according to \citet{Turunen1974} and \citet{Turunen1976}.

The effect of dislocation climb by both bulk and pipe diffusion on the vacancy concentration field is approximated by absorption (or emission) of point defects in proportion to the length and the distance travelled by a segment of a dislocation. Thus after each iteration of dislocation motion, the vacancy concentration is modified according to the area (in terms of lattice parameters) swept by each climbing dislocation segment. And the osmotic force $f_\mathrm{o}$ would be \citep{Hirth1982}
\begin{equation}
\frac{f_\mathrm{o}}{L} = -\frac{\mathrm{k_B} T b_{\mathrm{e}}}{v_{\mathrm{a}}}\ln \frac{c_\mathrm{v}}{c_0}
\label{eq:19}
\end{equation}
where $\mathrm{k_B}$ is the Boltzmann's constant; $b_\mathrm{e}$ is the edge component of Burgers vector; $v_\mathrm{a}$ is the atomic volume; and $c_0$ is the equilibrium concentration of vacancies, given by \citep{Lothe1960,Hirth1982}
\begin{equation}
c_0(\vectorsym{r}) = n_\mathrm{v} \mathrm{e}^{-\tfrac{E_\mathrm{f}-\sigma(\vectorsym{r}) \upDelta V}{\mathrm{k_B} T}}
\label{eq:c0}
\end{equation}
where $n_\mathrm{v}$ is the concentration of possible vacancy site; $\sigma(\vectorsym{r})$ is the stress experienced by the material at $\vectorsym{r}$ causing a change in volume $\upDelta V$; and $E_\mathrm{f}$ is the vacancy formation energy.

\subsubsection{Peierls forces}
The Peierls stress affects dislocation glide, acting as an energy barrier due to the varying misfit energy of the dislocation when it glides along a specific slip plane. This lattice frictional force varies with a periodicity closely related to nearest-neighbour interatomic distances within the material and depends on the slip system in which the dislocation lies. The temperature-dependent Peierls force $f_{\mathrm{p}}$ per unit length $L$, which includes the effect of thermal fluctuation at finite temperature, can be calculated by \citet{Chidambarrao1990} and \citet{Srinivasan2003}:
\begin{equation}
\frac{f_{\mathrm{p}}}{L} = 2\bar{\mu} |\vectorsym{b}| \cos{\phi} \left( \frac{1-\nu \cos^2 \alpha}{1- \nu}\omega \mathrm{e}^{\left( \tfrac{-2 {\uppi} d_{hkl} (1-\nu \cos^2 \alpha ) \omega}{(1- \nu) |\vectorsym{b}| } \right)} \right)
\label{eq:20}
\end{equation}
where $\phi$ is the angle between the film surface and the normal to the slip plane; $\alpha$ is the angle between the dislocation segment and Burgers vector; $\nu$ is the Poisson's ratio; $d_{hkl}$ is the interplanar spacing of the slip plane $(hkl)$ and $\omega$ is given by
\begin{equation}
\omega = \mathrm{e}^{\tfrac{4 \uppi^2n\mathrm{k_B}T}{5\bar{\mu}V}}
\label{eq:21}
\end{equation}
where $n$ is the number of atoms per unit cell; and $V$ is the volume of the unit cell.

\subsection{Dislocation mobility}
In the model, forces acting on dislocations produce dislocation movement via dislocation mobility functions. These functions convert calculated stresses into realistic simulated glide and climb velocities.

\subsubsection{Glide}
\begin{figure}[bht]
\centering\includegraphics{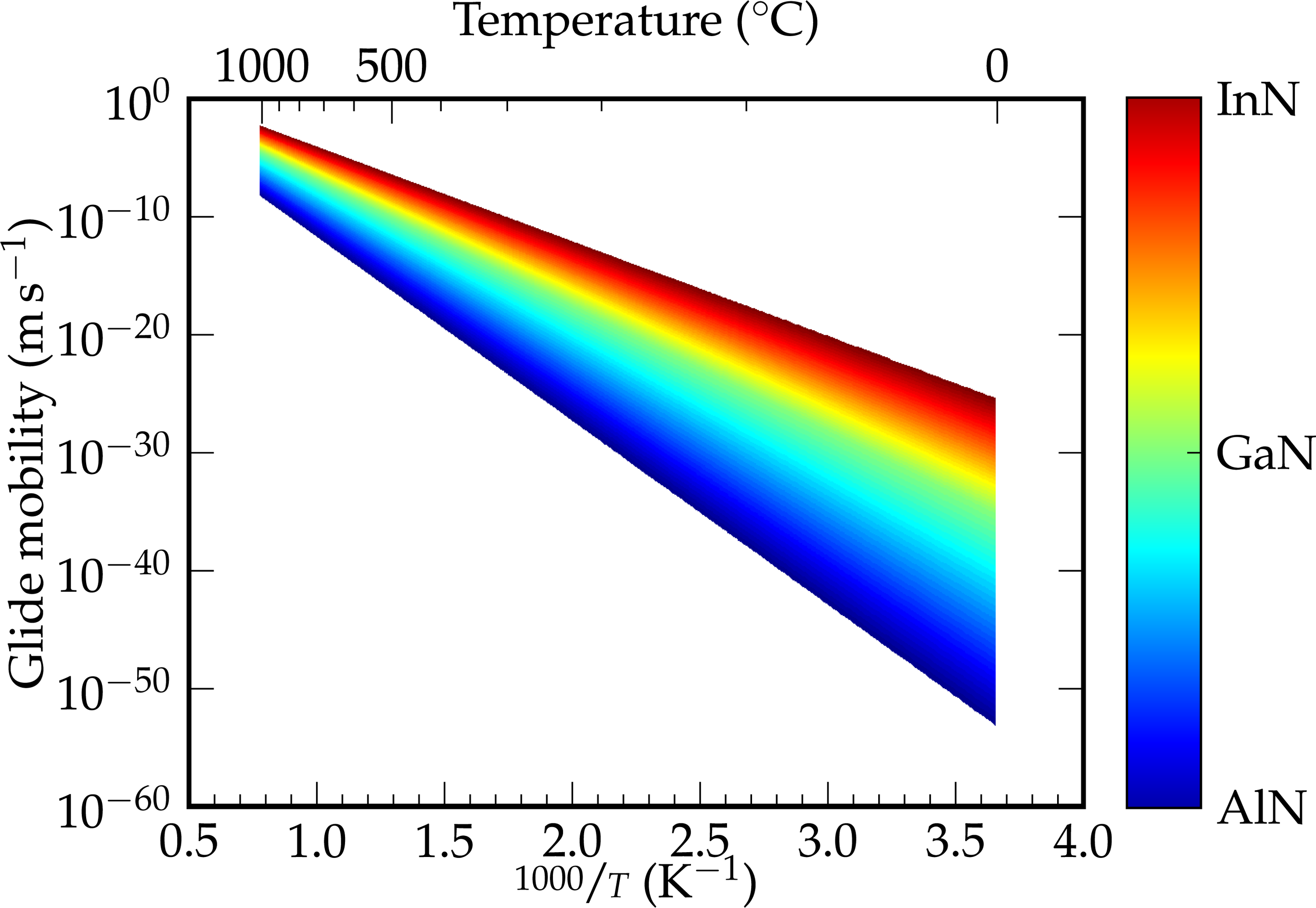}
\caption{Glide mobility versus temperature plot, calculated from the experimental data of \citet{Yonenaga2003a}, \citet{Yonenaga2003b,Yonenaga2009} and \citet{Sugiura1997}.}
\label{fig:Fig4}
\end{figure}
The simplest dislocation mobility function was proposed by \citet{Hirth1982} who used an arbitrary viscous drag coefficient $B$ to relate the force $f_\mathrm{g}$ per unit length $L$ and dislocation glide velocity $v_\mathrm{g}$. Mobility function $M$ (the inverse of $B$) has been used in some dislocation dynamics simulations in metals, but a suitable value of $M$ has not been proposed for nitrides \citep{Cai2004}. Although an accurate expression exists for $v_\mathrm{g}$ based on the classical theory of kink mobility \citep{Raabe1998}, several of its variables are unknown. Consequently, recent work has focused on determining $B$ accurately for each material of interest by fitting experimental data or molecular dynamics simulation, resulting in an empirical expression \citep{Fertig2009,Sugiura1997}. We adapt the expression in this work as: 
\begin{equation}
v_{\mathrm{g}} = \left[ v_0 \left( \frac{\tau_\mathrm{g}}{\tau_0} \right)^m \mathrm{e}^{- \tfrac{Q_{\mathrm{g}}}{\mathrm{k_B}T}} \right] \left( \vectorsym{n}_\mathrm{g} \times \vectorsym{\xi} \right)
\label{eq:22}
\end{equation}
where $v_\mathrm{g}$ is the glide velocity of a dislocation segment; $v_0$ is a constant; $\vectorsym{s}$ is the slip plane normal; $\tau_\mathrm{g}$ is the resolved shear stress acting on the dislocation segment; $\tau_0$ is a constant (1 MPa); and $Q_{\mathrm{g}}$ is the glide activation energy. From experimental data,  for GaN was estimated to be 2.0 eV \citep{Sugiura1997} or between 2.0 -- 2.7 eV \citep{Yonenaga2003a,Yonenaga2003b,Yonenaga2009}; whereas from molecular dynamics simulations, for GaN was estimated to be 1.6 eV, 1.6 eV and 4.7 eV for the basal, prismatic and pyramidal planes, respectively \citep{Weingarten2013}. The estimated glide mobility of III-nitrides is illustrated in Fig.~\ref{fig:Fig4}.

To account for dislocation glide in the model, the Peach-K\"{o}hler force acting on each dislocation segment is resolved in all the possible slip systems described by \citet{Srinivasan2003}. The resolved forces are then compared to the temperature-dependent Peierls forces of the corresponding slip systems. A slip system is designated as active if the net force on it is positive. The final slip direction is then selected according to the largest net glide force. 

\subsubsection{Climb}
It is more complicated to determine an accurate mobility function for dislocation climb. Climb can be driven by both hydrostatic and biaxial stresses and by osmotic effects. Climb is typically limited by the mobility of dislocation jogs \citep{Lothe1960} and can be enabled by bulk diffusion (i.e. diffusion of vacancies or interstitials through the bulk, towards or away from the dislocation) and/or by pipe diffusion (i.e. the preferential diffusion of vacancies or interstitials along the dislocation core). Pipe diffusion is likely to be highly relevant to III-nitride materials, because strong evidence of dislocation climb exists for GaN \citep{Moram2010} and AlN \citep{Fu2011} but both materials have very low vacancy and interstitial self-diffusion coefficients \citep{Laaksonen2009,Terentjevs2010} which are expected to lead to very low bulk climb velocities. 
Consequently, although a simple mobility function   has been used to model the climb velocity $v_\mathrm{c}$ \citep{Cai2004}, it cannot capture the complexity of climb processes occurring in real materials. Furthermore, although an expression has been derived for $v_\mathrm{c}$ based on the classical theory for bulk climb \citep{Raabe1998}, several of its variables are unknown and it cannot account for pipe-diffusion-controlled climb. A simpler expression for bulk climb has also been derived \citep{Clouet2011}, but it is not applicable to pipe-diffusion-controlled climb either. While climb mobility functions were also derived for both bulk and pipe diffusion of a single jog \citep{Lothe1960}, these assume that climb is driven only by osmotic forces and do not include any contribution from biaxial stress. 

In contrast, Turunen has derived a general equation of motion for a climbing, arbitrarily curved dislocation, taking into account both bulk diffusion \citep{Turunen1976} and pipe diffusion \citep{Turunen1974,Turunen1976} along the dislocation core. The climb velocity is described by:
\begin{equation}
v_{\mathrm{c}} = \left\{ \left( \frac{2 {\uppi} V^2 D c_0}{\mathrm{k_B} Tb_{\mathrm{e}}^2 \ln \sfrac{R}{\rho} } \right)f_{\mathrm{c}} - \left( \frac{2\nu_0 D \kappa a V^2}{\mathrm{k_B} Tb_{\mathrm{e}}^2} \mathrm{e}^{-\tfrac{Q_\mathrm{p}}{\mathrm{k_B} T}} \right) \left[ \frac{b_{\mathrm{s}}}{b_{\mathrm{e}}^2} \left( \vectorsym{b} \cdot \frac{\mathrm{d}\vectorsym{u}}{\mathrm{d}s}\frac{\mathrm{d}f_{\mathrm{c}}}{\mathrm{d}s}+\frac{\mathrm{d}^2 f_{\mathrm{c}}}{\mathrm{d}s^2} \right) \right] \right\} \frac{\vectorsym{b} \times \vectorsym{\xi}}{\left| \vectorsym{b} \times \vectorsym{\xi} \right|}
\label{eq:23}
\end{equation}
$V$ is the vacancy volume; $f_\mathrm{c}$ is the dislocation climb force; $\nu_0$ is the attempt frequency of atomic jumps; $Q_\mathrm{p}$ is the activation energy of pipe diffusion; $b_\mathrm{s}$ is the screw component of the Burgers vector; $\kappa$ is a numerical constant ({\raise.17ex\hbox{$\scriptstyle\sim$}}0.5); $a$ is the jump distance in the dislocation core; $R$ is an outer cutoff radius at $c(\vectorsym{r}) = c_0(\vectorsym{r})$; $\rho$ represents the dislocation core radius; and $\vectorsym{s}$ is a point along the dislocation line. The first term of the equation accounts for climb due to bulk diffusion, while the latter term accounts for climb due to pipe diffusion. The dislocation core is considered as a tunnel along the line direction made up of atomic sites where diffusive jumps could happen. From the climb force $f_\mathrm{c}$ and the activation energy required for a diffusion jump $Q_\mathrm{p}$, the effective number of diffused atoms causing climb of the dislocation segment can be related to the climb force, and thus $v_{\mathrm{c}}$ can be calculated from the net atomic flux diffusing along the dislocation core. This expression models both bulk and pipe diffusion and contains parameters that can be either measured or computed for III-nitrides, so therefore it is chosen for use in the simulations. 

We note that the $\dfrac{\mathrm{d}f_{\mathrm{c}}}{\mathrm{d}s}$ in the second term only exists when the dislocation line is not straight, i.e., $\dfrac{\mathrm{d}u}{\mathrm{d}s} \neq 0$: this was omitted in the original work based on the assumption that the dislocation has a low curvature \citep{Turunen1976}, which breaks down based on the simulations. Another point to note is that the Debye frequency, which is the theoretical maximum of the jump frequency, is more readily available from previous studies and is used as a sensible guess of the actual attempt frequency. It is calculated from the Debye temperature given in Table~\ref{tab:Tab2} by $\omega_0 = \frac{\mathrm{k_B}\omega_\mathrm{D}}{h}$. The Debye temperatures reported for GaN show a wide scatter between 500 K and 900 K and appropriate values have not yet been determined for all III-nitrides. However, the β values used in the fitting of the Varshni formula are closely associated with the Debye temperatures \citep{Roder2005,Teisseyre1994} and are available for all III-nitrides \citep{Vurgaftman2003}, so they have therefore been used in this work.

\subsection{Modelling of thin film growth}
Typically for nitride semiconductors, the most significant change in dislocation microstructures takes place during the growth of thin films and multilayers. Therefore, a simple growth model is introduced to the dislocation dynamics model, in which the film thickness can increase (or decrease, e.g. during etching) according to a predefined growth rate on each facet on the surface of the film, after which the stress field on dislocations is recalculated at each time step. This approach enables the response of dislocations to the change in film thickness to be modelled.

Complex film structures can then be 'grown' during the simulation using multiple 'growth' steps with step-specific time durations, the growth rate of the individual facet and material types. A simple growth mechanism is implemented such that the dislocation will choose to elongate along the direction of the substrate normal (usually $[0001]$) the surface facet that it is closest to, along other common dislocation directions for nitride films, $\left<11\bar{2}0\right>$ and $\left<10\bar{1}0\right>$, or along the original dislocation line direction, according to the lowest calculated strain energy that will be incurred.

\subsection{Choice of parameters used in the model}
Table~\ref{tab:Tab2} summarises the key material parameters used in the model, with references.

\begin{table}[hbt!]
\centering\resizebox*{\textwidth}{!}{
\begin{tabular}{ p{6cm} c c p{7cm}}
\toprule
Material parameters & \multicolumn{2}{c}{Value(s) chosen} & References\\ \midrule
&  $C_{11\mathrm{,AlN}} = 396$ GPa & $C_{12\mathrm{,AlN}} = 137$ GPa & \\
& $C_{13\mathrm{,AlN}} = 108$ GPa & $C_{33\mathrm{,AlN}} = 373$ GPa & \\
& $C_{44\mathrm{,AlN}} = 116$ GPa & & \\ \rule{0pt}{14pt}
& $C_{11\mathrm{,GaN}} = 367$ GPa & $C_{12\mathrm{,GaN}} = 135$ GPa & \\
& $C_{13\mathrm{,GaN}} = 103$ GPa & $C_{33\mathrm{,GaN}} = 405$ GPa & \\
& $C_{44\mathrm{,GaN}} = 95$ GPa & & \\ \rule{0pt}{14pt}
& $C_{11\mathrm{,InN}} = 223$ GPa & $C_{12\mathrm{,InN}} = 115$ GPa & \\
& $C_{13\mathrm{,InN}} = 92$ GPa & $C_{33\mathrm{,InN}} = 224$ GPa & \\
\multirow{-11}{6cm}{Elasticity tensor $\vectorsym{C}$} & $C_{44\mathrm{,InN}} = 48$ GPa & & \multirow{-11}{*}{\citet{Wright1997}} \\ \rule{0pt}{20pt}
\cellcolor[gray]{0.9}& \cellcolor[gray]{0.9}$a_{\mathrm{AlN}} = 3.111$ \angstrom & \cellcolor[gray]{0.9}$c_{\mathrm{AlN}} = 4.980$ \angstrom & \cellcolor[gray]{0.9}\\
\cellcolor[gray]{0.9}& \cellcolor[gray]{0.9}& \cellcolor[gray]{0.9}& \cellcolor[gray]{0.9}\\
\cellcolor[gray]{0.9}& \cellcolor[gray]{0.9}$a_{\mathrm{GaN}} = 3.189$ \angstrom & \cellcolor[gray]{0.9}$c_{\mathrm{GaN}} = 5.186$ \angstrom & \cellcolor[gray]{0.9}\\
\cellcolor[gray]{0.9}& \cellcolor[gray]{0.9} & \cellcolor[gray]{0.9}& \cellcolor[gray]{0.9}\\
\multirow{-5}{6cm}{\cellcolor[gray]{0.9}Lattice parameters $a$ and $c$} & \cellcolor[gray]{0.9}$a_{\mathrm{InN}} = 3.538$ \angstrom & \cellcolor[gray]{0.9}$c_{\mathrm{InN}} = 5.706$ \angstrom &   \multirow{-5}{7cm}{\cellcolor[gray]{0.9}\citet{Moram2009a}} \\
& \multicolumn{2}{c}{1462 K (AlN)} & \\
& \multicolumn{2}{c}{830 K (GaN)} & \\
\multirow{-4}{6cm}{Debye temperature $\theta_\mathrm{D}$} & \multicolumn{2}{c}{624 K (InN)} &  \multirow{-4}{7cm}{\citet{Roder2005,Teisseyre1994,Vurgaftman2003}} \\
\cellcolor[gray]{0.9} Dislocation core energy $E_\mathrm{core}$ & \multicolumn{2}{c}{\cellcolor[gray]{0.9}See Table~\ref{tab:Tab3}} & \cellcolor[gray]{0.9}\\
& a(AlN) & c(AlN) & \\
& a(GaN) & c(GaN) & \\
\multirow{-3}{6cm}{Temperature dependent thermal expansion function in the $\vectorsym{a}$ and $\vectorsym{c}$ directions} & a(InN) & c(InN) & \multirow{-3}{7cm}{\citet{Moram2009a,Reeber2000}} \\
\cellcolor[gray]{0.9}& \multicolumn{2}{c}{\cellcolor[gray]{0.9}$\mathrm{AlN} = 72.1 \mathrm{e}^{-\frac{4.7}{\mathrm{k_B}T}}$} & \cellcolor[gray]{0.9}\\
\cellcolor[gray]{0.9}& \multicolumn{2}{c}{\cellcolor[gray]{0.9}$\mathrm{GaN} = 43.0 \mathrm{e}^{-\frac{4.24}{\mathrm{k_B}T}}$} & \cellcolor[gray]{0.9}\\
\multirow{-3}{6cm}{\cellcolor[gray]{0.9}Diffusion coefficients\textsuperscript{*} for bulk diffusion $D_\mathrm{B}$} & \multicolumn{2}{c}{\cellcolor[gray]{0.9}$\mathrm{InN} = 39.8 \mathrm{e}^{-\frac{2.1}{\mathrm{k_B}T}}$} & \multirow{-3}{7cm}{\cellcolor[gray]{0.9}Extrapolated from \citet{Ambacher1998}} \\
& \multicolumn{2}{c}{$Q_\mathrm{g}(\mathrm{AlN}) = 3.1$ eV} & \\ 
& \multicolumn{2}{c}{$Q_\mathrm{g}(\mathrm{GaN}) = 2.1$ eV} & \\ 
\multirow{-3}{6cm}{Glide activation energy $Q_\mathrm{g}$} & \multicolumn{2}{c}{$Q_\mathrm{g}(\mathrm{InN}) = 1.2$ eV} & \multirow{-3}{7cm}{Estimated from \citet{Sugiura1997,Yonenaga2009}}\\ 
\cellcolor[gray]{0.9}& \multicolumn{2}{c}{\cellcolor[gray]{0.9}$Q_\mathrm{p}(\mathrm{AlN}) = 3.4$ eV} & \cellcolor[gray]{0.9}\\ 
\multirow{-2}{6cm}{\cellcolor[gray]{0.9}Climb activation energy, for pipe diffusion $Q_\mathrm{p}$} & \multicolumn{2}{c}{\cellcolor[gray]{0.9}$Q_\mathrm{p}(\mathrm{GaN}) = 3.7$ eV} & \multirow{-2}{7cm}{\cellcolor[gray]{0.9}\citet{Fu2011,Moram2010}}\\
& \multicolumn{2}{c}{$V_\mathrm{Al} = 10.44$ \angstrom\textsuperscript{3}} & \\ 
& \multicolumn{2}{c}{$V_\mathrm{Ga} = 11.42$ \angstrom\textsuperscript{3}} & \\ 
\multirow{-3}{6cm}{Atomic volume $V_{\mathrm{A}}$} & \multicolumn{2}{c}{$V_\mathrm{In} = 15.46$ \angstrom\textsuperscript{3}} & \multirow{-3}{7cm}{Estimated from lattice parameters \citet{Moram2009a} } \\ \bottomrule
\multicolumn{4}{c}{\footnotesize\textsuperscript{*}includes the vacancy or interstitial formation energy, the diffusion coefficient pre-exponential factor and the self-diffusion enthalpy}\\
\end{tabular}}
\caption{Parameters used in simulations of thin films and multilayers of AlN, GaN, InN and their alloys. N.B. The value for $Q_\mathrm{p}$ of InN is currently unknown as the relevant experimental studies are not available in the literature.}
\label{tab:Tab2}
\end{table} 

\subsubsection{Core radii and core energies}
\begin{table}
\centering\resizebox*{\textwidth}{!}{
\begin{tabular}{ c M{1.7cm} m{2.3cm} m{2cm} m{4cm} m{3cm} }
\toprule
Material & Burgers vector $\vectorsym{b}$ & Core radius $\rho$ (\angstrom) & $E_\mathrm{core}$ (eV \angstrom\textsuperscript{-1}) & Method & Reference \\ \midrule
\cellcolor[gray]{0.9} & \aplusc- type & 7.2 & 3.12 & MSW potential & \cite{Belabbas2007} \\ 
\cellcolor[gray]{0.9} & \cellcolor[gray]{0.9} & \cellcolor[gray]{0.9} 7.0 & \cellcolor[gray]{0.9} 1.57 & \cellcolor[gray]{0.9} Multiscale DFT with SW potential & \cellcolor[gray]{0.9} \cite{Lymperakis2005} \\ 
\multirow{-3}{*}[10pt]{\cellcolor[gray]{0.9}GaN} & \cellcolor[gray]{0.9} & 6.0 & 1.61 & \multirow{3}{4cm}{Many body interatomic Tersoff potential} & \multirow{3}{3cm}{\cite{Kioseoglou2009}} \\ 
AlN & \cellcolor[gray]{0.9} & 8.3 & 1.71 & & \\ 
InN & \multirow{-4}{*}[8pt]{\cellcolor[gray]{0.9}$\vectorsym{a}$-type} & 5.4 & 1.66 & & \\
\bottomrule
\end{tabular}}
\caption{Core radii and energies for dislocations in III-nitrides, including $\vectorsym{a}$-type dislocations with 5/7-atom ring cores, and \aplusc-type dislocations with 5/6-atom ring cores from atomistic simulations.}
\label{tab:Tab3}
\end{table}

The three types of pure dislocations in III-nitrides are summarised in Table~\ref{tab:Tab2}. However, approximately 99\% of the dislocations found in heteroepitaxial III-nitride structures on sapphire are either $\vectorsym{a}$-type or \aplusc-type \citep{Moram2009b}, apart from at the early stages of film growth where different ratios of different dislocation types can occur. Different core structures are possible \citep{Lymperakis2004}, but the commonest are the 5/7-atom ring structure for $\vectorsym{a}$-type dislocations and the 5/6-atom ring structure for the \aplusc-type dislocations \citep{Rhode2013}. Therefore, we have used reliable literature values for the core radii and the core energies of dislocations in III-nitrides with these core structures, as listed in Table~\ref{tab:Tab3}.

\subsubsection{Activation energies for dislocation climb}
\begin{figure}[bht]
\centering\includegraphics{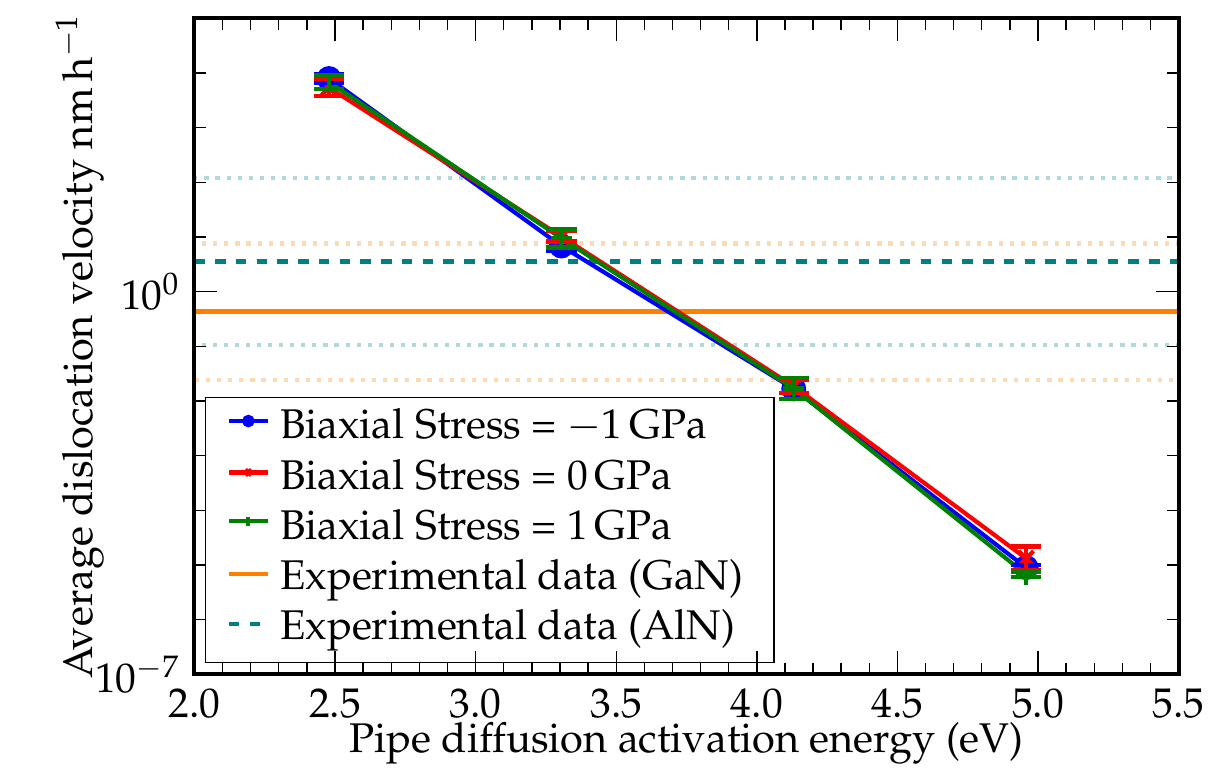}
\caption{Average dislocation velocities plotted versus pipe diffusion activation energy, subjected to different external biaxial stresses. Experimental data for GaN and AlN are extracted from the results of \citet{Moram2010} (solid orange line) and \citet{Fu2011} (dashed greenish blue line), respectively. Their error ranges are represented by the dotted lines with light orange colour for GaN and light greenish blue colour for AlN.}
\label{fig:Fig5}
\end{figure}
Even though the precise mechanism of climb is not known in the III-nitrides, the activation energy of the rate-limiting step of the climb process can still be identified by comparison to experimental data. The model was used to simulate the growth of GaN- and AlN-on-sapphire films according to the conditions reported by \citet{Moram2010} and \citet{Fu2011} respectively and average dislocation velocities were obtained. These results were then compared to the average dislocation velocities obtained from the experimental results reported by \citet{Moram2010} and \citet{Fu2011} respectively. The upper and lower error bounds are obtained by assuming no and maximum dislocation inclination responsible for the displacement of dislocation. Fig.~\ref{fig:Fig5} shows the average dislocation velocities obtained from simulations and plotted against different pipe diffusion activation energies for GaN films (each simulation was repeated 10 times; results for AlN were very similar). A small change in $Q\mathrm{p}$ alters the dislocation climb velocities by orders of magnitude, while the simulated climb velocity is much less sensitive to the other simulation parameters. This is as expected considering Eqn.~\ref{eq:23} for climb mobility, where the climb velocity is exponentially dependent on the activation energy. From Fig.~\ref{fig:Fig5}, $Q\mathrm{p}$ for AlN and GaN are thus found to be approximately $3.4 \pm 0.45$ eV and $3.7 \pm 0.40$ eV respectively. However, the average climb dislocation velocity estimated from experiment shows that dislocations climb faster in AlN than in GaN, so AlN has a lower pipe diffusion activation energy. The difference in pipe diffusion activation energies between GaN and AlN may be due to relative size effects: the atomic radius of Al is approximately 6\% smaller than that of Ga (135 pm and 143 pm, respectively) \citep{Rhode2013}, whereas the a lattice parameters and consequently the internal diameters of the dislocation cores are only 2.5\% smaller for AlN than for GaN ($a_\mathrm{AlN} = 3.111$ \angstrom and $a_\mathrm{GaN} = 3.189$ \angstrom) \citep{Moram2009a}. In the simulations, the rate of diffusion is proportional to $\mathrm{e}^{-\frac{Q_0}{\mathrm{k_B}T}}$ and to the cross sectional area of the dislocation pipe $\uppi r_\mathrm{c}$, a decrease in activation energy with increasing relative core radius $r_\mathrm{c}$ is expected. 

\section{Results}
The simulations were validated by comparison to transmission electron microscopy data of an AlN epilayer on a sapphire substrate and of an \AlGaNSample\ film on an AlN-on-sapphire substrate.
For the AlN film on sapphire, the simulation was set up using a simulation cell with a lateral size of 169 nm $\times$ 169 nm containing 10 dislocations with randomly assigned Burgers vector directions and with types assigned randomly according to the proportions found experimentally. In this case, the simulations included three $\vectorsym{a}$-type dislocations, six $\vectorsym{c}$-type dislocations and one \aplusc-type dislocation, as these ratios are in proportion to the amounts of different types of dislocations found experimentally at the very early stages of film growth \citep{Fu2011} (note that the proportion of $\vectorsym{a}$-type dislocations rises greatly as the film thickness increases, because $\vectorsym{c}$-type dislocations tend to annihilate each other easily during the initial stages of film growth \citep{Fu2011}. All other settings were based on experimental growth parameters, including an epilayer growth temperature of 1403 K and a growth rate of 1.3 \umph. Here, a free surface was defined at the bottom of the AlN film to simulate the effects of the disordered nucleation layer found experimentally between AlN films and sapphire substrates. The effects of the experimentally verified sapphire substrate miscut of 0.25\degree towards the $[11\bar{2}0]$ direction were tested by performing one simulation including the miscut and one simulation without it.

For the \AlGaNSample\ film on AlN, the simulation was set up  using a simulation cell with a lateral size of 396 nm $\times$ 396 nm containing 10 dislocations with randomly assigned Burgers vector directions and with types assigned randomly according to the proportions found experimentally, in this case nine a-type dislocations and one \aplusc-type dislocation. All other settings were based on experimental growth parameters, including an epilayer growth temperature of 1382 K and a growth rate of 2 \umph. The interface is a continuum boundary and the simulations were initiated assuming that the pre-existing dislocations in the AlN layer (beneath the AlGaN film) were initially straight and aligned along $[0001]$.

\begin{figure}[bht]
\centering\includegraphics{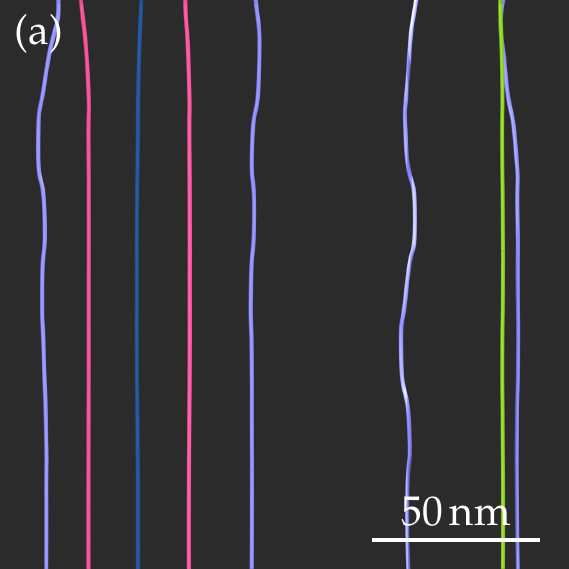}
\includegraphics{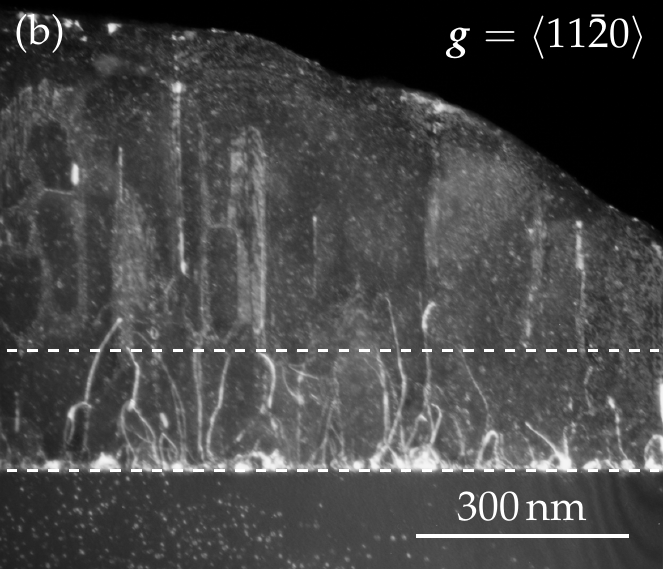}
\caption{(a) Simulation of an AlN film grown on sapphire with a substrate miscut of 0.25\degree towards $[11\bar{2}0]$; (b) cross-sectional weak-beam dark field transmission electron micrograph of a 1 \um\ thick AlN film grown on a sapphire substrate, with a substrate miscut of $0.25 \pm 0.10\degreem$ towards $[11\bar{2}0]$. The micrograph was taken in the $\vectorsym{g}(5\vectorsym{g})$ condition with $\vectorsym{g} = 11\bar{2}0$, revealing $\vectorsym{a}$-type and \aplusc-type dislocations. The plane of the TEM foil and the plane on which the contents of the simulation cell are projected is $\left(10\bar{1}0\right)$. Different colours are used for each dislocation in the simulation, for ease of identification. The upper and lower dashed lines in (b) correspond to the vertical boundaries of the simulation as in (a).}
\label{fig:Fig6}
\end{figure}

Fig.~\ref{fig:Fig6} shows the simulation results and the corresponding experimentally determined microstructure for the AlN film on sapphire. The colours of the dislocations in the figure represent their corresponding Burgers vectors ($b_\mathrm{x}$, $b_\mathrm{y}$ and $b_\mathrm{z}$ are mapped to red, green and blue colours, respectively). No dislocation bending is predicted in simulations of the growth of AlN on sapphire without any miscut: all dislocations remain oriented parallel to $[0001]$. However, when a substrate miscut is included in the simulations, the model reproduces accurately the effects of the miscut on the microstructure of the AlN films. For example, the substrate miscut of 0.25\degree towards $[11\bar{2}0]$ results in the onset of dislocation bending away from $[0001]$ at an AlN thickness of approximately 150 nm, consistent with experimental data. These data indicate that the substrate miscut has an important role to play in inducing dislocation bending, which is well known to result in an increase in dislocation annihilation and reduction with increasing epilayer thickness in III-nitride films.

\begin{figure}[bht!]
\centering
\includegraphics[height=6cm]{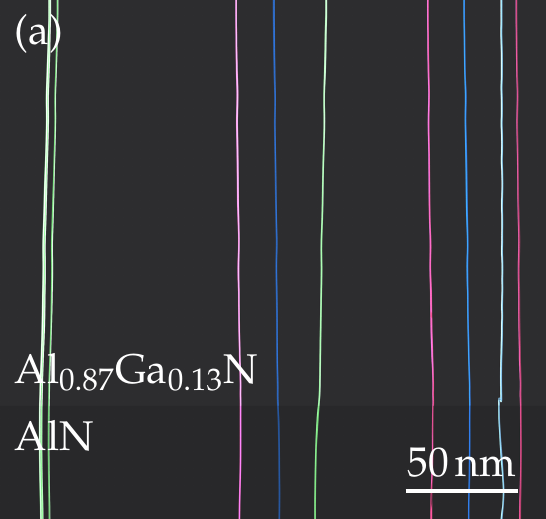}
\includegraphics[height=6cm]{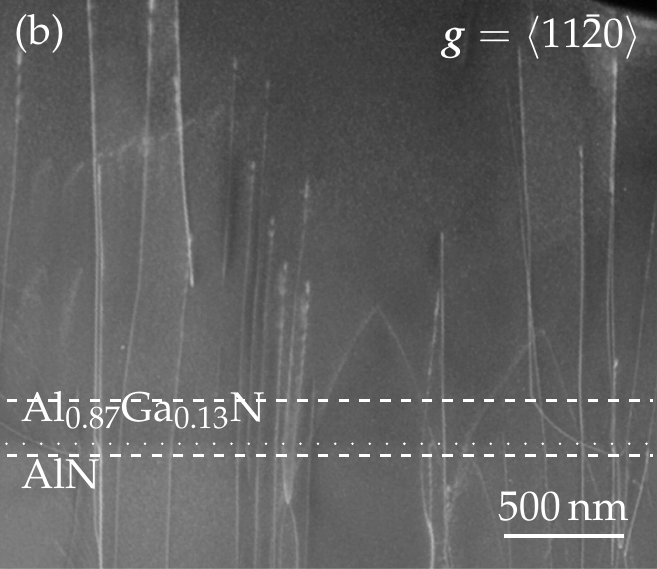}
\caption{(a) Simulation of an \AlGaNSample film grown on sapphire with a substrate miscut of 0.25\degree towards $[11\bar{2}0]$; (b) cross-sectional weak-beam dark field transmission electron micrograph of an \AlGaNSample film on an AlN buffer layer. The micrograph was taken in the $\vectorsym{g}(5\vectorsym{g})$ condition with $\vectorsym{g} = 11\bar{2}0$, revealing $\vectorsym{a}$-type and \aplusc-type dislocations. The plane of the TEM foil and the plane on which the contents of the simulation cell are projected is $\left(10\bar{1}0\right)$. Different colours are used for each dislocation in the simulation, for ease of identification. The upper and lower dashed lines in (b) correspond to the vertical boundaries of the simulation as in (a). The dotted line represents the heterointerface of \AlGaNSample/AlN.}
\label{fig:Fig7}
\end{figure}

Fig.~\ref{fig:Fig7} shows the simulation results and the corresponding experimentally determined microstructure for the \AlGaNSample/AlN heterostructure. The average (projected) dislocation bending angle away from [0001] in the TEM specimen was $6.7\degreem \pm 1.4\degreem$, whereas the projected dislocation bending angle away from $[0001]$ from the simulations was $6.7\degreem \pm 0.6\degreem$, i.e. the same, within the standard error. Minor differences may relate to the fact that the image force from the AlN/sapphire interface was not included in this simulation to minimise computational expense, as the microstructure was expected to be dominated by the effects of misfit stresses at the \AlGaNSample/AlN interface. The simulations match the experimental microstructure accurately and are consistent with the low experimentally observed strain relaxation of 6\%. These results are in contrast to widely cited equilibrium critical thickness calculations for the \AlGaNx/AlN system (equivalent to the inverse of the \AlGaNx/GaN calculations), which predict incorrectly that stress relaxation should occur by dislocation glide at an \AlGaNSample\ thickness of just 80 nm \citep{Holec2008}. The discrepancy occurs because the equilibrium critical thickness calculations assume that dislocation glide is the only stress relief mechanism that can act in these heterostructures, whereas our model includes the effects of dislocation climb, of image stresses and of the influence of neighbouring dislocations (each with their own strain field). This result indicates that dislocation climb plays an important role in stress relaxation in AlGaN-based heterostructures. The effects of dislocation climb on critical thicknesses for stress relaxation will be explored further in a subsequent publication.

\section{Discussions}
Table~\ref{tab:Tab4} shows a list of 3D discrete dislocation dynamics simulation packages, including this model. A comparison of our model considering the crucial features for simulating dislocation dynamics in thin film nitride semiconductors is also presented. Most of the newer packages have incorporated elastic anisotropy, although only microMegas has explicitly stated and demonstrated the ability to handle hexagonal crystals \citep{Monnet2004}. Most of the packages have been used to study strained epitaxial films with a heterointerface terminated by a free surface, although only MDDP \citep{Zbib2002} and PDD \citep{Ghoniem2005} have demonstrated the ability to handle multilayered structures (i.e. those containing more than two layers). However, most of these studies focus on the ‘channelling stress', that is, the thickness-dependent critical stress driving a threading dislocation to introduce a misfit segment. Realistic misfit stresses are much less studied and data appear only for metallic heterointerfaces \citep{Groh2003}, rather than for semiconductors. Therefore, the key advantage of the present approach is the successful implementation and combination of the following features:

\begin{enumerate}[i.]
\item	The ability to handle both climb and glide, including climb driven by both osmotic and mechanical stresses and enabled by both bulk and pipe diffusion.
\item	Full elastic anisotropy for materials with hexagonal symmetry.
\item	Efficient calculations for thin film multilayers and free surfaces, including irregular geometries (e.g. islanded layers), using a superposition method \citep{Tan2006}.
\item	Ability to simulate dislocation dynamics under different thin film growth conditions, including the effects of changes in growth temperature between different layers in a heterostructure.
\end{enumerate}
\afterpage{
\begin{landscape}
\noindent
\begin{table}[h!]
\centering
\noindent
\resizebox*{!}{0.72\textheight}{
\begin{tabular}{ m{2.5cm} m{2.5cm} m{3cm} m{1.62cm} m{1.62cm} m{1.62cm} m{1.62cm} m{1.62cm} m{2.74cm} m{5cm} }
\toprule
Simulation code & Discretization & Stress calculation & Aniso-tropy & hcp & Free surface & Multi-layer & Misfit forces & Climb & References\\ \midrule
This model & Straight line & Willis-Steeds-Lothe & Yes & Yes & Yes & Yes & Yes & Yes (bulk and pipe) & -\\
\rowcolor[gray]{0.9}
K-D & Edge-screw & Brown & & Yes (line tension) & Yes & Bilayer & Yes & & \cite{Brown1964,Groh2003}\\
MDDP & Straight line & Simple dislocation bend & & & Yes & Yes & Yes & & \cite{Zbib2001,Zbib2002,Akasheh2007}\\
\rowcolor[gray]{0.9}
microMega & Straight line & Modified de Wit & Yes & Yes & Yes & & & & \cite{Devincre1995,Monnet2004,Devincre2011}\\
ParaDiS & Straight line & Willis-Steeds-Lothe & Yes & & Yes & & & Yes (glide-like mobility law) & \cite{Bulatov2004,Cai2004,Meijie2006,Arsenlis2007,Yin2010}\\
\rowcolor[gray]{0.9}
PARANOID & Tracking points & Modified Brown & & & Yes & Bilayer & Yes & & \cite{Schwarz1996,Schwarz1999,Liu2005}\\
PDD & Cubic spline & Han and Ghoniem & & Yes & Yes & Yes (planar films) & Yes & Yes (bulk) & \cite{Ghoniem2000,Han2003,Ghoniem2005}\\
\rowcolor[gray]{0.9}
Tridis & Edge-screw & Modified de Wit & & & Yes & & & Yes (bulk) & \cite{Devincre1995,Hartmaier1999,Mordehai2008}\\
VGA & Tracking points & Modified Brown & & No & Yes & Bilayer & Yes & & \cite{Schwarz1999,vonBlanckenhagen2001,vonBlanckenhagen2004}\\
\rowcolor[gray]{0.9}
W & Straight line & Brown & & & Yes & & & & \cite{Brown1964,Weygand2002} \\
\bottomrule
\end{tabular}}
\caption{A list of 3D discrete dislocation dynamics simulation packages available currently, including features that are either published or explicitly described in the corresponding model's description in the accompanying manual or on the relevant host website.}
\label{tab:Tab4}
\end{table}
\end{landscape}
}
\section{Conclusions}
In this work we have described PANIC, a new model for discrete dislocation dynamics simulations which uses an adaptive multi-scale meshing approach combined with input from atomistic simulations to reproduce mesoscale dislocation behaviour. This code can accommodate both hexagonal and cubic materials and includes the full effects of elastic anisotropy. Multilayer structures can be simulated, including the effects of multiple free surfaces and interfaces. Misfit forces and thermal stresses can be included, while arbitrary geometries (both planar and non-planar) can also be simulated. This code has been validated for the case of technologically important AlN and AlGaN-based thin film heterostructures, for which simulation code inputs are well known. This is the first dislocation dynamics simulation to accurately model dislocation behaviour within semiconductor heterostructures: it is therefore anticipated that the code will facilitate the design and development of devices including defect-containing heterostructures, especially for the challenging III-nitride materials system.

\section*{Acknowledgement}
WYF acknowledges funding through a Croucher Foundation Scholarship. MAM acknowledges funding through a Royal Society University Research Fellowship and through EPSRC grant no. EP/J015792. This work was performed using the Darwin Supercomputer of the University of Cambridge High Performance Computing Service, provided by Dell Inc. using Strategic Research Infrastructure Funding from the Higher Education Funding Council for England and funding from the Science and Technology Facilities Council.

\bibliographystyle{elsarticle-harv} 
\bibliography{PANICLib}

\begin{thebibliography}{112}
\expandafter\ifx\csname natexlab\endcsname\relax\def\natexlab#1{#1}\fi
\expandafter\ifx\csname url\endcsname\relax
  \def\url#1{\texttt{#1}}\fi
\expandafter\ifx\csname urlprefix\endcsname\relax\def\urlprefix{URL }\fi

\bibitem[{Akasheh et~al.(2007)Akasheh, Zbib, Hirth, Hoagland, and
  Misra}]{Akasheh2007}
Akasheh, F., Zbib, H.~M., Hirth, J.~P., Hoagland, R.~G., Misra, A., 2007.
  {Dislocation dynamics analysis of dislocation intersections in nanoscale
  metallic multilayered composites}. Journal of Applied Physics 101~(8),
  084314.

\bibitem[{Amano et~al.(1999)Amano, Iwaya, Hayashi, Kashima, Nitta, Wetzel, and
  Akasaki}]{Amano1999}
Amano, H., Iwaya, M., Hayashi, N., Kashima, T., Nitta, S., Wetzel, C., Akasaki,
  I., 1999. {Control of Dislocations and Stress in AlGaN on Sapphire Using a
  Low Temperature Interlayer}. physica status solidi (b) 216~(1), 683--689.

\bibitem[{Ambacher et~al.(1998)Ambacher, Freudenberg, Dimitrov, Angerer, and
  Stutzmann}]{Ambacher1998}
Ambacher, O., Freudenberg, F., Dimitrov, R., Angerer, H., Stutzmann, M., 1998.
  {Nitrogen Effusion and Self-Diffusion in Ga$_{14}$N/Ga$_{15}$N Isotope
  Heterostructures}. Japanese Journal of Applied Physics 37, 2416.

\bibitem[{Arsenlis et~al.(2007)Arsenlis, Cai, Tang, Rhee, Oppelstrup, Hommes,
  Pierce, and Bulatov}]{Arsenlis2007}
Arsenlis, A., Cai, W., Tang, M., Rhee, M., Oppelstrup, T., Hommes, G., Pierce,
  T.~G., Bulatov, V.~V., 2007. {Enabling strain hardening simulations with
  dislocation dynamics}. Modelling and Simulation in Materials Science and
  Engineering 15~(6), 553.

\bibitem[{Asaro and Barnett(1974)}]{Asaro1974}
Asaro, R.~J., Barnett, D.~M., 1974. {On some three-dimensional dislocation
  problems in anisotropic media}. Journal of Physics F: Metal Physics 4~(5),
  L103.

\bibitem[{Ayas et~al.(2014)Ayas, van Dommelen, and Deshpande}]{Ayas2014}
Ayas, C., van Dommelen, J., Deshpande, V., 2014. {Climb-enabled discrete
  dislocation plasticity}. Journal of the Mechanics and Physics of Solids 62,
  113--136.

\bibitem[{Belabbas et~al.(2007)Belabbas, B\'{e}r\'{e}, Chen, Petit, Belkhir,
  Ruterana, and Nouet}]{Belabbas2007}
Belabbas, I., B\'{e}r\'{e}, A., Chen, J., Petit, S., Belkhir, M.~A., Ruterana,
  P., Nouet, G., 2007. {Atomistic modeling of the ($a+c$)-mixed dislocation
  core in wurtzite GaN}. Physical Review B 75~(11), 115201.

\bibitem[{Brown(1964)}]{Brown1964}
Brown, L.~M., 1964. {The self-stress of dislocations and the shape of extended
  nodes}. Philosophical Magazine 10~(105), 441--466.

\bibitem[{Brown(1967)}]{Brown1967}
Brown, L.~M., 1967. {A proof of lothe's theorem}. Philosophical Magazine
  15~(134), 363--370.

\bibitem[{Bulatov et~al.(1998)Bulatov, Abraham, Kubin, Devincre, and
  Yip}]{Bulatov1998}
Bulatov, V., Abraham, F.~F., Kubin, L., Devincre, B., Yip, S., 1998.
  {Connecting atomistic and mesoscale simulations of crystal plasticity}.
  Nature 391~(6668), 669--672.

\bibitem[{Bulatov et~al.(2004)Bulatov, Cai, Fier, Hiratani, Hommes, Pierce,
  Tang, Rhee, Yates, and Arsenlis}]{Bulatov2004}
Bulatov, V., Cai, W., Fier, J., Hiratani, M., Hommes, G., Pierce, T., Tang, M.,
  Rhee, M., Yates, K., Arsenlis, T., 2004. {Scalable Line Dynamics in ParaDiS}.
  SC '04. IEEE Computer Society.

\bibitem[{Cai et~al.(2006)Cai, Arsenlis, Weinberger, and Bulatov}]{Cai2006}
Cai, W., Arsenlis, A., Weinberger, C.~R., Bulatov, V.~V., 2006. {A non-singular
  continuum theory of dislocations}. Journal of the Mechanics and Physics of
  Solids 54~(3), 561--587.

\bibitem[{Cai and Bulatov(2004)}]{Cai2004}
Cai, W., Bulatov, V.~V., 2004. {Mobility laws in dislocation dynamics
  simulations}. Materials Science and Engineering: A 387-389, 277--281.

\bibitem[{Chan et~al.(2009)Chan, Hou, Tseng, Chen, Chien, Hsiao, Lee, Tsai, and
  Chen}]{Chan2009}
Chan, C.-H., Hou, C.-H., Tseng, S.-Z., Chen, T.-J., Chien, H.-T., Hsiao, F.-L.,
  Lee, C.-C., Tsai, Y.-L., Chen, C.-C., 2009. {Improved output power of
  GaN-based light-emitting diodes grown on a nanopatterned sapphire substrate}.
  Applied Physics Letters 95~(1), 011110.

\bibitem[{Chang et~al.(2010)Chang, Moram, McAleese, Kappers, and
  Humphreys}]{Chang2010}
Chang, T.~Y., Moram, M.~A., McAleese, C., Kappers, M.~J., Humphreys, C.~J.,
  2010. {Inclined dislocation arrays in AlGaN/AlGaN quantum well structures
  emitting at 290 nm}. Journal of Applied Physics 108~(12), 123522.

\bibitem[{Cherns et~al.(2008)Cherns, {McAleese}, Kappers, and
  Humphreys}]{Cherns2008}
Cherns, P.~D., {McAleese}, C., Kappers, M.~J., Humphreys, C.~J., 2008. {Strain
  Relaxation in an AlGaN/GaN Quantum Well System}. In: Cullis, A.~G., Midgley,
  P.~A. (Eds.), Microscopy of Semiconducting Materials 2007. Vol. 120 of
  Springer Proceedings in Physics. Springer Netherlands, pp. 25--28.

\bibitem[{Chidambarrao et~al.(1990)Chidambarrao, Srinivasan, Cunningham, and
  Murthy}]{Chidambarrao1990}
Chidambarrao, D., Srinivasan, G.~R., Cunningham, B., Murthy, C.~S., 1990.
  {Effects of Peierls barrier and epithreading dislocation orientation on the
  critical thickness in heteroepitaxial structures}. Applied Physics Letters
  57~(10), 1001.

\bibitem[{Clouet(2011)}]{Clouet2011}
Clouet, E., 2011. {Predicting dislocation climb: Classical modeling versus
  atomistic simulations}. Physical Review B 84~(9), 092106.

\bibitem[{Costa et~al.(2005)Costa, Datta, Kappers, Vickers, and
  Humphreys}]{Costa2005}
Costa, P.~M., Datta, R., Kappers, M.~J., Vickers, M.~E., Humphreys, C.~J.,
  2005. {Misfit dislocations in green-emitting InGaN/GaN quantum well
  structures}. MRS Proceedings 892, FF25--01.

\bibitem[{Cottrell(1963)}]{Cottrell1963}
Cottrell, A.~H., 1963. {Dislocations and plastic flow in crystals}. Clarendon
  Press.

\bibitem[{Cserti et~al.(1992)Cserti, Khantha, Vitek, and Pope}]{Cserti1992}
Cserti, J., Khantha, M., Vitek, V., Pope, D., 1992. {An atomistic study of the
  dislocation core structures and mechanical behavior of a model D0$_{19}$
  alloy}. Materials Science and Engineering: A 152~(1-2), 95--102.

\bibitem[{Dadgar et~al.(2000)Dadgar, Bl\"{a}sing, Diez, Alam, Heuken, and
  Krost}]{Dadgar2000}
Dadgar, A., Bl\"{a}sing, J., Diez, A., Alam, A., Heuken, M., Krost, A., 2000.
  {Metalorganic Chemical Vapor Phase Epitaxy of Crack-Free GaN on Si (111)
  Exceeding 1\um\ in Thickness}. Japanese Journal of Applied Physics 39~(Part
  2, No. 11B), L1183--L1185.

\bibitem[{Deng et~al.(2008)Deng, El-Azab, and Larson}]{Deng2008}
Deng, J., El-Azab, A., Larson, B., 2008. {On the elastic boundary value problem
  of dislocations in bounded crystals}. Philosophical Magazine 88~(30-32),
  3527--3548.

\bibitem[{Devincre(1995)}]{Devincre1995}
Devincre, B., 1995. {Three dimensional stress field expressions for straight
  dislocation segments}. Solid State Communications 93~(11), 875--878.

\bibitem[{Devincre et~al.(2011)Devincre, Madec, Monnet, Queyreau, Gatti, and
  Kubin}]{Devincre2011}
Devincre, B., Madec, R., Monnet, G., Queyreau, S., Gatti, R., Kubin, L., 2011.
  {Modeling crystal plasticity with dislocation dynamics simulations: The
  'microMegas' code}.

\bibitem[{Dular et~al.(1998)Dular, Geuzaine, Henrotte, and Legros}]{Dular1998}
Dular, P., Geuzaine, C., Henrotte, F., Legros, W., 1998. {A general environment
  for the treatment of discrete problems and its application to the finite
  element method}. Ieee Transactions on Magnetics 34~(5), 3395--3398.

\bibitem[{Fang and Wang(2000)}]{Fang2000}
Fang, Q.~F., Wang, R., 2000. {Atomistic simulation of the atomic structure and
  diffusion within the core region of an edge dislocation in aluminum}.
  Physical Review B 62~(14), 9317--9324.

\bibitem[{Fertig and Baker(2009)}]{Fertig2009}
Fertig, R.~S., Baker, S.~P., 2009. {Simulation of dislocations and strength in
  thin films: A review}. Progress in Materials Science 54~(6), 874--908.

\bibitem[{Fitzgerald and Aubry(2010)}]{Fitzgerald2010}
Fitzgerald, S.~P., Aubry, S., 2010. {Self-force on dislocation segments in
  anisotropic crystals}. Journal of Physics: Condensed Matter 22~(29), 295403.

\bibitem[{Fivel et~al.(1998)Fivel, Robertson, Canova, and
  Boulanger}]{Fivel1998}
Fivel, M., Robertson, C., Canova, G., Boulanger, L., 1998. {Three-dimensional
  modeling of indent-induced plastic zone at a mesoscale}. Acta Materialia
  46~(17), 6183--6194.

\bibitem[{Forghani et~al.(2012)Forghani, Schade, Schwarz, Lipski, Klein,
  Kaiser, and Scholz}]{Forghani2012}
Forghani, K., Schade, L., Schwarz, U.~T., Lipski, F., Klein, O., Kaiser, U.,
  Scholz, F., 2012. {Strain and defects in Si-doped (Al)GaN epitaxial layers}.
  Journal of Applied Physics 112~(9), 093102.

\bibitem[{Freund and Suresh(2003)}]{Freund2003}
Freund, L.~B., Suresh, S., 2003. {Thin Film Materials: Stress, Defect Formation
  and Surface Evolution}. Cambridge University Press.

\bibitem[{Fu et~al.(2011)Fu, Kappers, Zhang, Humphreys, and Moram}]{Fu2011}
Fu, W.~Y., Kappers, M.~J., Zhang, Y., Humphreys, C.~J., Moram, M.~A., 2011.
  {Dislocation Climb in $c$-Plane AlN Films}. Applied Physics Express 4~(6),
  65503.

\bibitem[{Furitsch et~al.(2006)Furitsch, Avramescu, Eichler, Engl, Leber,
  Miler, Rumbolz, Br\"{u}derl, Strau$\mkern-1.35mu\upbeta$, Lell, and
  H\"{a}rle}]{Furitsch2006}
Furitsch, M., Avramescu, A., Eichler, C., Engl, K., Leber, A., Miler, A.,
  Rumbolz, C., Br\"{u}derl, G., Strau$\mkern-1.35mu\upbeta$, U., Lell, A.,
  H\"{a}rle, V., 2006. {Comparison of degradation mechanisms of blue-violet
  laser diodes grown on SiC and GaN substrates}. physica status solidi (a)
  203~(7), 1797--1801.

\bibitem[{Geuzaine and Remacle(2009)}]{Geuzaine2009}
Geuzaine, C., Remacle, J.-F., 2009. {Gmsh: A 3-D finite element mesh generator
  with built-in pre- and post-processing facilities}. International Journal for
  Numerical Methods in Engineering 79~(11), 1309--1331.

\bibitem[{Ghoniem and Han(2005)}]{Ghoniem2005}
Ghoniem, N.~M., Han, X., 2005. {Dislocation motion in anisotropic multilayer
  materials}. Philosophical Magazine 85~(24), 2809--2830.

\bibitem[{Ghoniem et~al.(2000)Ghoniem, Tong, and Sun}]{Ghoniem2000}
Ghoniem, N.~M., Tong, S.-H., Sun, L.~Z., 2000. {Parametric dislocation
  dynamics: A thermodynamics-based approach to investigations of mesoscopic
  plastic deformation}. Physical Review B 61~(2), 913--927.

\bibitem[{Groh et~al.(2003)Groh, Devincre, Kubin, Roos, Feyel, and
  Chaboche}]{Groh2003}
Groh, S., Devincre, B., Kubin, L.~P., Roos, A., Feyel, F., Chaboche, J.-L.,
  2003. {Dislocations and elastic anisotropy in heteroepitaxial metallic thin
  films}. Philosophical Magazine Letters 83~(5), 303--313.

\bibitem[{Haeberlen et~al.(2010)Haeberlen, Zhu, McAleese, Kappers, and
  Humphreys}]{Haeberlen2010}
Haeberlen, M., Zhu, D., McAleese, C., Kappers, M.~J., Humphreys, C.~J., 2010.
  {Dislocation reduction in MOVPE grown GaN layers on (111)Si using SiN$_x$ and
  AlGaN layers}. Journal of Physics: Conference Series 209~(1), 012017.

\bibitem[{Han et~al.(2001)Han, Waldrip, Lee, Figiel, Hearne, Petersen, and
  Myers}]{Han2001}
Han, J., Waldrip, K.~E., Lee, S.~R., Figiel, J.~J., Hearne, S.~J., Petersen,
  G.~A., Myers, S.~M., 2001. {Control and elimination of cracking of AlGaN
  using low-temperature AlGaN interlayers}. Applied Physics Letters 78~(1), 67.

\bibitem[{Han et~al.(2003)Han, Ghoniem, and Wang}]{Han2003}
Han, X., Ghoniem, N.~M., Wang, Z., 2003. {Parametric dislocation dynamics of
  anisotropic crystals}. Philosophical Magazine 83~(31-34), 3705--3721.

\bibitem[{Hartmaier et~al.(1999)Hartmaier, Fivel, Canova, and
  Gumbsch}]{Hartmaier1999}
Hartmaier, A., Fivel, M.~C., Canova, G.~R., Gumbsch, P., 1999. {Image stresses
  in a free-standing thin film}. Modelling and Simulation in Materials Science
  and Engineering 7~(5), 781.

\bibitem[{Head(1953{\natexlab{a}})}]{Head1953b}
Head, A., 1953{\natexlab{a}}. {X. The interaction of dislocations and
  boundaries}. Philosophical Magazine Series 7 44~(348), 92--94.

\bibitem[{Head(1953{\natexlab{b}})}]{Head1953a}
Head, A.~K., 1953{\natexlab{b}}. {Edge Dislocations in Inhomogeneous Media}.
  Proceedings of the Physical Society B 66~(9), 793.

\bibitem[{Hirsch et~al.(1956)Hirsch, Horne, and Whelan}]{Hirsch1956}
Hirsch, P.~B., Horne, R.~W., Whelan, M.~J., 1956. {LXVIII. Direct observations
  of the arrangement and motion of dislocations in aluminium}. Philosophical
  Magazine 1~(7), 677--684.

\bibitem[{Hirth and Lothe(1982)}]{Hirth1982}
Hirth, J.~P., Lothe, J., 1982. {Theory of dislocations}, 2nd Edition.
  John-Wiley \& Sons.

\bibitem[{Holec et~al.(2008)Holec, Zhang, Rao, Kappers, McAleese, and
  Humphreys}]{Holec2008}
Holec, D., Zhang, Y., Rao, D. V.~S., Kappers, M.~J., McAleese, C., Humphreys,
  C.~J., 2008. {Equilibrium critical thickness for misfit dislocations in
  III-nitrides}. Journal of Applied Physics 104~(12), 123514.

\bibitem[{Hsu et~al.(2012)Hsu, Hardy, Young, Romanov, DenBaars, Nakamura, and
  Speck}]{Hsu2012}
Hsu, P.~S., Hardy, M.~T., Young, E.~C., Romanov, A.~E., DenBaars, S.~P.,
  Nakamura, S., Speck, J.~S., 2012. {Stress relaxation and critical thickness
  for misfit dislocation formation in $\left(10\bar{1}0\right)$ and
  $\left(30\bar{3}\bar{1}\right)$ InGaN/GaN heteroepitaxy}. Applied Physics
  Letters 100~(17), 171917.

\bibitem[{Hsu et~al.(2011)Hsu, Young, Romanov, Fujito, DenBaars, Nakamura, and
  Speck}]{Hsu2011}
Hsu, P.~S., Young, E.~C., Romanov, A.~E., Fujito, K., DenBaars, S.~P.,
  Nakamura, S., Speck, J.~S., 2011. {Misfit dislocation formation via
  pre-existing threading dislocation glide in $\left(11\bar{2}2\right)$
  semipolar heteroepitaxy}. Applied Physics Letters 99~(8), 081912.

\bibitem[{Kaun et~al.(2011)Kaun, Wong, Dasgupta, Choi, Chung, Mishra, and
  Speck}]{Kaun2011}
Kaun, S.~W., Wong, M.~H., Dasgupta, S., Choi, S., Chung, R., Mishra, U.~K.,
  Speck, J.~S., 2011. {Effects of Threading Dislocation Density on the Gate
  Leakage of AlGaN/GaN Heterostructures for High Electron Mobility
  Transistors}. Applied Physics Express 4~(2), 024101.

\bibitem[{Khan et~al.(2008)Khan, Balakrishnan, and Katona}]{Khan2008}
Khan, A., Balakrishnan, K., Katona, T., 2008. {Ultraviolet light-emitting
  diodes based on group three nitrides}. Nature Photonics 2~(2), 77--84.

\bibitem[{Kioseoglou et~al.(2009)Kioseoglou, Komninou, and
  Karakostas}]{Kioseoglou2009}
Kioseoglou, J., Komninou, P., Karakostas, T., 2009. {Core models of $a$-edge
  threading dislocations in wurtzite III(Al,Ga,In)-nitrides}. physica status
  solidi (a) 206~(8), 1931--1935.

\bibitem[{Krost and Dadgar(2002)}]{Krost2002}
Krost, A., Dadgar, A., 2002. {GaN-Based Devices on Si}. physica status solidi
  (a) 194~(2), 361--375.

\bibitem[{Laaksonen et~al.(2009)Laaksonen, Ganchenkova, and
  Nieminen}]{Laaksonen2009}
Laaksonen, K., Ganchenkova, M.~G., Nieminen, R.~M., 2009. {Vacancies in
  wurtzite GaN and AlN}. Journal of Physics: Condensed Matter 21~(1), 015803.

\bibitem[{Liu and Schwarz(2005)}]{Liu2005}
Liu, X.~H., Schwarz, K.~W., 2005. {Modelling of dislocations intersecting a
  free surface}. Modelling and Simulation in Materials Science and Engineering
  13~(8), 1233.

\bibitem[{Lothe(1960)}]{Lothe1960}
Lothe, J., 1960. {Theory of dislocation climb in metals}. Journal of Applied
  Physics 31~(6), 1077--1087.

\bibitem[{Lothe(1967)}]{Lothe1967}
Lothe, J., 1967. {Dislocation bends in anisotropic media}. Philosophical
  Magazine 15~(134), 353--362.

\bibitem[{Lothe et~al.(1982)Lothe, Indenbom, and Chamrov}]{Lothe1982}
Lothe, J., Indenbom, V.~L., Chamrov, V.~A., 1982. {Elastic field and self-force
  of dislocations emerging at the free surfaces of an anisotropic halfspace}.
  physica status solidi (b) 111~(2), 671--677.

\bibitem[{Lymperakis(2005)}]{Lymperakis2005}
Lymperakis, L., 2005. {Ab-initio Based Multiscale Calculations of Extended
  Defects in and on Group III-nitrides}. Ph.D. thesis, Universit\"{a}t
  Paderborn.

\bibitem[{Lymperakis et~al.(2004)Lymperakis, Neugebauer, Albrecht, Remmele, and
  Strunk}]{Lymperakis2004}
Lymperakis, L., Neugebauer, J., Albrecht, M., Remmele, T., Strunk, H.~P., 2004.
  {Strain Induced Deep Electronic States around Threading Dislocations in GaN}.
  Physical Review Letters 93~(19), 196401.

\bibitem[{Mathews and Fink(2006)}]{Mathews2006}
Mathews, J.~H., Fink, K.~D., 2006. {Numerical Methods Using MATLAB}. Pearson
  Education, Limited.

\bibitem[{Matsumoto and Nishimura(1998)}]{Matsumoto1998}
Matsumoto, M., Nishimura, T., 1998. {Mersenne twister: a 623-dimensionally
  equidistributed uniform pseudo-random number generator}. ACM Trans. Model.
  Comput. Simul. 8~(1), 3--30.

\bibitem[{McAleese et~al.(2004)McAleese, Kappers, Rayment, Cherns, and
  Humphreys}]{McAleese2004}
McAleese, C., Kappers, M.~J., Rayment, F. D.~G., Cherns, P., Humphreys, C.~J.,
  2004. {Strain effects of AlN interlayers for MOVPE growth of crack-free AlGaN
  and AlN/GaN multilayers on GaN}. Journal of Crystal Growth 272~(1-4),
  475--480.

\bibitem[{Meijie et~al.(2006)Meijie, Cai, Xu, Bulatov, and Tang}]{Meijie2006}
Meijie, T., Cai, W., Xu, G., Bulatov, V.~V., Tang, M., 2006. {A hybrid method
  for computing forces on curved dislocations intersecting free surfaces in
  three-dimensional dislocation dynamics}. Modelling and Simulation in
  Materials Science and Engineering 14~(7), 1139.

\bibitem[{Monnet et~al.(2004)Monnet, Devincre, and Kubin}]{Monnet2004}
Monnet, G., Devincre, B., Kubin, L., 2004. {Dislocation study of prismatic slip
  systems and their interactions in hexagonal close packed metals: application
  to zirconium}. Acta Materialia 52~(14), 4317--4328.

\bibitem[{Moram et~al.(2009)Moram, Ghedia, Rao, Barnard, Zhang, Kappers, and
  Humphreys}]{Moram2009b}
Moram, M.~A., Ghedia, C.~S., Rao, D. V.~S., Barnard, J.~S., Zhang, Y., Kappers,
  M.~J., Humphreys, C.~J., 2009. {On the origin of threading dislocations in
  GaN films}. Journal of Applied Physics 106~(7), 073513.

\bibitem[{Moram et~al.(2011{\natexlab{a}})Moram, Kappers, Massabuau, Oliver,
  and Humphreys}]{Moram2011a}
Moram, M.~A., Kappers, M.~J., Massabuau, F., Oliver, R.~A., Humphreys, C.~J.,
  2011{\natexlab{a}}. {Response to "Comment on 'The effects of Si doping on
  dislocation movement and tensile stress in GaN films'" [J. Appl. Phys. 109,
  073509 (2011)]}. Journal of Applied Physics 110~(9), 096102.

\bibitem[{Moram et~al.(2011{\natexlab{b}})Moram, Kappers, Massabuau, Oliver,
  and Humphreys}]{Moram2011b}
Moram, M.~A., Kappers, M.~J., Massabuau, F., Oliver, R.~A., Humphreys, C.~J.,
  2011{\natexlab{b}}. {The effects of Si doping on dislocation movement and
  tensile stress in GaN films}. Journal of Applied Physics 109~(7), 073509.

\bibitem[{Moram et~al.(2010)Moram, Sadler, Haberlen, Kappers, and
  Humphreys}]{Moram2010}
Moram, M.~A., Sadler, T.~C., Haberlen, M., Kappers, M.~J., Humphreys, C.~J.,
  2010. {Dislocation movement in GaN films}. Applied Physics Letters 97~(26),
  261903--261907.

\bibitem[{Moram and Vickers(2009)}]{Moram2009a}
Moram, M.~A., Vickers, M.~E., 2009. {X-ray diffraction of III-nitrides}.
  Reports on Progress in Physics 72~(3), 36502.

\bibitem[{Mordehai et~al.(2008)Mordehai, Clouet, Fivel, and
  Verdier}]{Mordehai2008}
Mordehai, D., Clouet, E., Fivel, M., Verdier, M., 2008. {Introducing
  dislocation climb by bulk diffusion in discrete dislocation dynamics}.
  Philosophical Magazine 88~(6), 899--925.

\bibitem[{Mukai et~al.(2006)Mukai, Morita, Yamamoto, Akaishi, Matoba, Yasutomo,
  Kasai, Sano, and Nagahama}]{Mukai2006}
Mukai, T., Morita, D., Yamamoto, M., Akaishi, K., Matoba, K., Yasutomo, K.,
  Kasai, Y., Sano, M., Nagahama, S.-i., 2006. {Investigation of
  optical-output-power degradation in 365-nm UV-LEDs}. physica status solidi
  (c) 3~(6), 2211--2214.

\bibitem[{Olmsted et~al.(2005)Olmsted, Jr, Curtin, and Clifton}]{Olmsted2005}
Olmsted, D.~L., Jr, L. G.~H., Curtin, W.~A., Clifton, R.~J., 2005. {Atomistic
  simulations of dislocation mobility in Al, Ni and Al/Mg alloys}. Modelling
  and Simulation in Materials Science and Engineering 13~(3), 371.

\bibitem[{Orowan(1934{\natexlab{a}})}]{Orowan1934a}
Orowan, E., 1934{\natexlab{a}}. {Zur Kristallplastizit\"{a}}t. i. Zeitschrift
  f\"{u}r Physik 89~(9-10), 605--613.

\bibitem[{Orowan(1934{\natexlab{b}})}]{Orowan1934b}
Orowan, E., 1934{\natexlab{b}}. {Zur Kristallplastizit\"{a}}t. ii. Zeitschrift
  f\"{u}r Physik 89~(9-10), 614--633.

\bibitem[{Orowan(1934{\natexlab{c}})}]{Orowan1934c}
Orowan, E., 1934{\natexlab{c}}. {Zur Kristallplastizit\"{a}}t. iii. Zeitschrift
  f\"{u}r Physik 89~(9-10), 634--659.

\bibitem[{Polanyi(1934)}]{Polanyi1934}
Polanyi, M., 1934. {\"{U}}ber eine art gitterst\"{o}rung, die einen kristall
  plastisch machen k\"{o}nnte. Zeitschrift f\"{u}r Physik 89~(9-10), 660--664.

\bibitem[{Raabe(1998)}]{Raabe1998}
Raabe, D., 1998. {Computational materials science: the simulation of materials
  microstructures and properties}. Wiley-VCH.

\bibitem[{Reeber and Wang(2000)}]{Reeber2000}
Reeber, R.~R., Wang, K., 2000. {Lattice parameters and thermal expansion of
  important semiconductors and their substrates} 622, T6.35.

\bibitem[{Rhode et~al.(2013)Rhode, Horton, Kappers, Zhang, Humphreys, Dusane,
  Sahonta, and Moram}]{Rhode2013}
Rhode, S.~K., Horton, M.~K., Kappers, M.~J., Zhang, S., Humphreys, C.~J.,
  Dusane, R.~O., Sahonta, S.~L., Moram, M.~A., 2013. {Mg Doping Affects
  Dislocation Core Structures in GaN}. Physical Review Letters 111~(2), 025502.

\bibitem[{Roder et~al.(2005)Roder, Einfeldt, Figge, and Hommel}]{Roder2005}
Roder, C., Einfeldt, S., Figge, S., Hommel, D., 2005. {Temperature dependence
  of the thermal expansion of GaN}. Physical Review B 72~(8), 085218.

\bibitem[{Schwarz(1999)}]{Schwarz1999}
Schwarz, K.~W., 1999. {Simulation of dislocations on the mesoscopic scale. I.
  Methods and examples}. Journal of Applied Physics 85~(1), 108--119.

\bibitem[{Schwarz and Tersoff(1996)}]{Schwarz1996}
Schwarz, K.~W., Tersoff, J., 1996. {Interaction of threading and misfit
  dislocations in a strained epitaxial layer}. Applied Physics Letters 69~(9),
  1220--1222.

\bibitem[{Srinivasan et~al.(2003)Srinivasan, Geng, Liu, Ponce, Narukawa, and
  Tanaka}]{Srinivasan2003}
Srinivasan, S., Geng, L., Liu, R., Ponce, F.~A., Narukawa, Y., Tanaka, S.,
  2003. {Slip systems and misfit dislocations in InGaN epilayers}. Applied
  Physics Letters 83~(25), 5187--5189.

\bibitem[{Steeds(1973)}]{Steeds1973}
Steeds, J.~W., 1973. {Introduction to anisotropic elasticity theory of
  dislocations}. Clarendon Press.

\bibitem[{Sugiura(1997)}]{Sugiura1997}
Sugiura, L., 1997. {Dislocation motion in GaN light-emitting devices and its
  effect on device lifetime}. Journal of Applied Physics 81~(4), 1633--1638.

\bibitem[{Tan and Sun(2006)}]{Tan2006}
Tan, E.~H., Sun, L.~Z., 2006. {Stress field due to a dislocation loop in a
  heterogeneous thin film-substrate system}. Modelling and Simulation in
  Materials Science and Engineering 14~(6), 993.

\bibitem[{Tapajna et~al.(2011)Tapajna, Kaun, Wong, Gao, Palacios, Mishra,
  Speck, and Kuball}]{Tapajna2011}
Tapajna, M., Kaun, S.~W., Wong, M.~H., Gao, F., Palacios, T., Mishra, U.~K.,
  Speck, J.~S., Kuball, M., 2011. {Influence of threading dislocation density
  on early degradation in AlGaN/GaN high electron mobility transistors}.
  Applied Physics Letters 99~(22), 223501.

\bibitem[{Taylor(1934)}]{Taylor1934}
Taylor, G.~I., 1934. {The Mechanism of Plastic Deformation of Crystals. Part I.
  Theoretical}. Proceedings of the Royal Society of London A 145~(855),
  362--387.

\bibitem[{Teisseyre et~al.(1994)Teisseyre, Perlin, Suski, Grzegory, Porowski,
  Jun, Pietraszko, and Moustakas}]{Teisseyre1994}
Teisseyre, H., Perlin, P., Suski, T., Grzegory, I., Porowski, S., Jun, J.,
  Pietraszko, A., Moustakas, T.~D., 1994. {Temperature dependence of the energy
  gap in GaN bulk single crystals and epitaxial layer}. Journal of Applied
  Physics 76~(4), 2429.

\bibitem[{Terentjevs et~al.(2010)Terentjevs, Catellani, and
  Cicero}]{Terentjevs2010}
Terentjevs, A., Catellani, A., Cicero, G., 2010. {Nitrogen vacancies at InN
  $\left(1\bar{1}00\right)$ surfaces: A theoretical study}. Applied Physics
  Letters 96~(17), 171901.

\bibitem[{Turunen(1976)}]{Turunen1976}
Turunen, M.~J., 1976. {A general equation of motion for dislocation climb}.
  Acta Metallurgica 24~(5), 463--467.

\bibitem[{Turunen and Lindroos(1974)}]{Turunen1974}
Turunen, M.~J., Lindroos, V.~K., 1974. {Model for dislocation climb by a pipe
  diffusion mechanism}. Philosophical Magazine 29~(4), 701--708.

\bibitem[{Van~der Giessen and Needleman(1995)}]{VanderGiessen1995}
Van~der Giessen, E., Needleman, A., 1995. {Discrete dislocation plasticity: a
  simple planar model}. Modelling and Simulation in Materials Science and
  Engineering 3~(5), 689.

\bibitem[{von Blanckenhagen et~al.(2004)von Blanckenhagen, Arzt, and
  Gumbsch}]{vonBlanckenhagen2004}
von Blanckenhagen, B., Arzt, E., Gumbsch, P., 2004. {Discrete dislocation
  simulation of plastic deformation in metal thin films}. Acta Materialia
  52~(3), 773--784.

\bibitem[{von Blanckenhagen et~al.(2001)von Blanckenhagen, Gumbsch, and
  Arzt}]{vonBlanckenhagen2001}
von Blanckenhagen, B., Gumbsch, P., Arzt, E., 2001. {Dislocation sources in
  discrete dislocation simulations of thin-film plasticity and the Hall-Petch
  relation}. Modelling and Simulation in Materials Science and Engineering
  9~(3), 157.

\bibitem[{Vurgaftman and Meyer(2003)}]{Vurgaftman2003}
Vurgaftman, I., Meyer, J.~R., 2003. {Band parameters for nitrogen-containing
  semiconductors}. Journal of Applied Physics 94~(6), 3675--3696.

\bibitem[{Wang and Reeber(2001)}]{Wang2001}
Wang, K., Reeber, R.~R., 2001. {Thermal expansion and elastic properties of
  InN}. Applied Physics Letters 79~(11), 1602--1604.

\bibitem[{Weinberger et~al.(2009)Weinberger, Aubry, Lee, and
  Cai}]{Weinberger2009}
Weinberger, C.~R., Aubry, S., Lee, S.-W., Cai, W., 2009. {Dislocation dynamics
  simulations in a cylinder}. IOP Conference Series: Materials Science and
  Engineering 3~(1), 012007.

\bibitem[{Weingarten and Chung(2013)}]{Weingarten2013}
Weingarten, N.~S., Chung, P.~W., 2013. {$a$-Type edge dislocation mobility in
  wurtzite GaN using molecular dynamics}. Scripta Materialia 82~(4), 311--314.

\bibitem[{Weygand et~al.(2002)Weygand, Friedman, Giessen, and
  Needleman}]{Weygand2002}
Weygand, D., Friedman, L.~H., Giessen, E. V.~d., Needleman, A., 2002. {Aspects
  of boundary-value problem solutions with three-dimensional dislocation
  dynamics}. Modelling and Simulation in Materials Science and Engineering
  10~(4), 437.

\bibitem[{Willis(1970)}]{Willis1970}
Willis, J.~R., 1970. {Stress fields produced by dislocations in anisotropic
  media}. Philosophical Magazine 21~(173), 931--949.

\bibitem[{Wright(1997)}]{Wright1997}
Wright, A.~F., 1997. {Elastic properties of zinc-blende and wurtzite AlN, GaN,
  and InN}. Journal of Applied Physics 82~(6), 2833--2839.

\bibitem[{Yasin et~al.(2001)Yasin, Zbib, and Khaleel}]{Yasin2001}
Yasin, H., Zbib, H.~M., Khaleel, M.~A., 2001. {Size and boundary effects in
  discrete dislocation dynamics: coupling with continuum finite element}.
  Materials Science and Engineering: A 309-310, 294--299.

\bibitem[{Yin et~al.(2010)Yin, Barnett, and Cai}]{Yin2010}
Yin, J., Barnett, D.~M., Cai, W., 2010. {Efficient computation of forces on
  dislocation segments in anisotropic elasticity}. Modelling and Simulation in
  Materials Science and Engineering 18~(4), 045013.

\bibitem[{Yoffe(1961)}]{Yoffe1961}
Yoffe, E.~H., 1961. {A dislocation at a free surface}. Philosophical Magazine
  6~(69), 1147--1155.

\bibitem[{Yonenaga(2003)}]{Yonenaga2003a}
Yonenaga, I., 2003. {High-temperature strength of bulk single crystals of III-V
  nitrides}. Journal of Materials Science: Materials in Electronics 14~(5),
  279--281.

\bibitem[{Yonenaga et~al.(2003)Yonenaga, Itoh, and Goto}]{Yonenaga2003b}
Yonenaga, I., Itoh, S., Goto, T., 2003. {Dislocation mobility and
  photoluminescence of plastically deformed GaN}. Physica B: Condensed Matter
  340-342, 484--487.

\bibitem[{Yonenaga et~al.(2009)Yonenaga, Ohno, Taishi, and
  Tokumoto}]{Yonenaga2009}
Yonenaga, I., Ohno, Y., Taishi, T., Tokumoto, Y., 2009. {Recent knowledge of
  strength and dislocation mobility in wide band-gap semiconductors}. Physica
  B: Condensed Matter 404~(23-24), 4999--5001.

\bibitem[{Zbib and Di\'{a}z de~la Rubia(2002)}]{Zbib2002}
Zbib, H.~M., Di\'{a}z de~la Rubia, T., 2002. {A Multiscale Model of
  Plasticity}. International Journal of Plasticity 18~(9), 1133--1163.

\bibitem[{Zbib et~al.(2001)Zbib, Di\'{a}z de~la Rubia, and Bulatov}]{Zbib2001}
Zbib, H.~M., Di\'{a}z de~la Rubia, T., Bulatov, V.~V., 2001. {A Multiscale
  Model of Plasticity Based on Discrete Dislocation Dynamics}. Journal of
  Engineering Materials and Technology 124~(1), 78--87.

\bibitem[{Zhu et~al.(2013)Zhu, Wallis, and Humphreys}]{Zhu2013}
Zhu, D., Wallis, D.~J., Humphreys, C.~J., 2013. {Prospects of III-nitride
  optoelectronics grown on Si}. Reports on Progress in Physics 76~(10), 106501.

\end{thebibliography}

\end{document}